\documentclass[journal]{IEEEtran}
\usepackage{cite}
\usepackage{rotating} 
\usepackage{amsmath}
\usepackage{url}
\usepackage{tikz}

\newcommand\submittedtext{%
  \footnotesize This work has been submitted to the IEEE for possible publication. Copyright may be transferred without notice, after which this version may no longer be accessible.}

\newcommand\submittednotice{%
\begin{tikzpicture}[remember picture,overlay]
\node[anchor=south,yshift=10pt] at (current page.south) {\fbox{\parbox{\dimexpr0.65\textwidth-\fboxsep-\fboxrule\relax}{\submittedtext}}};
\end{tikzpicture}%
}

\ifCLASSINFOpdf
\else
\fi
\begin{document}
%
\title{Open-source End-to-End Digital Beamforming System Modeling}

\author{Jose~Guajardo, Ali~Niknejad}

\maketitle
\section{Introduction}
\submittednotice
The goal of this paper is to introduce an open-source MATLAB-based behavioral hardware  model of a general digital beamforming system. More specifically, it models an end-to-end uplink between an arbitrary number of user elements (UEs) and an arbitrarily large base station (BS) with and without a strong interferer. This paper also presents and validates an equation-based model for the effects of interference on thermal and quantization noise. The paper is organized as follows: firstly, Section  \ref{section:ChDBFModelOverview} provides background and gives an overview of the behavioral model, detailing the model architecture, variable definitions and introducing model outputs and heuristics. Then, Section \ref{section:ChDBFResults} presents various results of the behavioral model, such as results for determining the maximum tolerable interferer strength and minimum ADC resolution for various digital beamforming systems. Additionally, results on thermal noise effects, one-bit ADC topologies and user count effects are presented. Lastly, Section \ref{section:ChDBFConclusions} concludes with a summary of contributions and a discussion of future directions. Appendix \ref{section:appendix} provides the repository location and usage instructions.

\section{Background and Model Overview}
\label{section:ChDBFModelOverview}
\subsection{Background: Complex Envelope Representation}
In a typical communication system, a modulated signal is often represented as a sinusoidal carrier wave with varying amplitude and phase. This signal can be expressed in terms of two orthogonal components:
\begin{itemize}
    \item \textbf{In-phase (I)}: This component is aligned with the cosine of the carrier frequency.
    \item \textbf{Quadrature (Q)}: This component is aligned with the sine of the carrier frequency.
\end{itemize}

The overall modulated signal is a combination of these two components, where the in-phase signal modulates the cosine wave and the quadrature signal modulates the sine wave. The \( I \) and \( Q \) components carry the information about the amplitude and phase of the signal.

Consider a real passband signal \( s(t) \) that is modulated by a carrier frequency \( f_c \). It can be represented as:

\[
s(t) = I(t) \cdot \cos(2\pi f_c t) - Q(t) \cdot \sin(2\pi f_c t)
\]

Here:
\begin{itemize}
    \item \( I(t) \) is the in-phase component, modulating the cosine wave.
    \item \( Q(t) \) is the quadrature component, modulating the sine wave.
\end{itemize}

The minus sign indicates that the \( Q \) component is 90 degrees out of phase with the \( I \) component.

To simplify the representation and processing, we can use the concept of a complex envelope. The complex envelope \( c(t) \) of the signal is defined as:

\[
c(t) = I(t) + jQ(t)
\]

Using this complex envelope, the real passband signal \( s(t) \) can be expressed as:

\[
s(t) = \text{Re}\{ c(t) \cdot e^{j2\pi f_c t} \}
\]

Where \( e^{j2\pi f_c t} = \cos(2\pi f_c t) + j\sin(2\pi f_c t) \) represents the carrier wave.

Expanding this, we get:

\[
s(t) = \text{Re}\{ \left( I(t) + jQ(t) \right) \cdot \left( \cos(2\pi f_c t) + j\sin(2\pi f_c t) \right) \}
\]

This simplifies to:

\[
s(t) = I(t) \cdot \cos(2\pi f_c t) - Q(t) \cdot \sin(2\pi f_c t)
\]

Which is the same as the original signal representation, showing that the complex envelope \( c(t) \) captures both the \( I \) and \( Q \) components in a single, complex-valued function.

The complex envelope allows us to work with lower-frequency, baseband signals while retaining all the essential information about the modulated signal. This abstraction simplifies many tasks in signal processing, including modulation, filtering, and demodulation, without the need to deal with the high-frequency carrier directly.

The behavioral model presented in this paper utilizes the complex envelope to operate on lower frequency signals, saving simulation time. Since the high-frequency carrier is not simulated, this abstraction does not allow the model to include some effects related to the physical transmission of RF signals, such as non-linearities in RF components.

\begin{figure}
    \centering
    \includegraphics[width=1\linewidth]{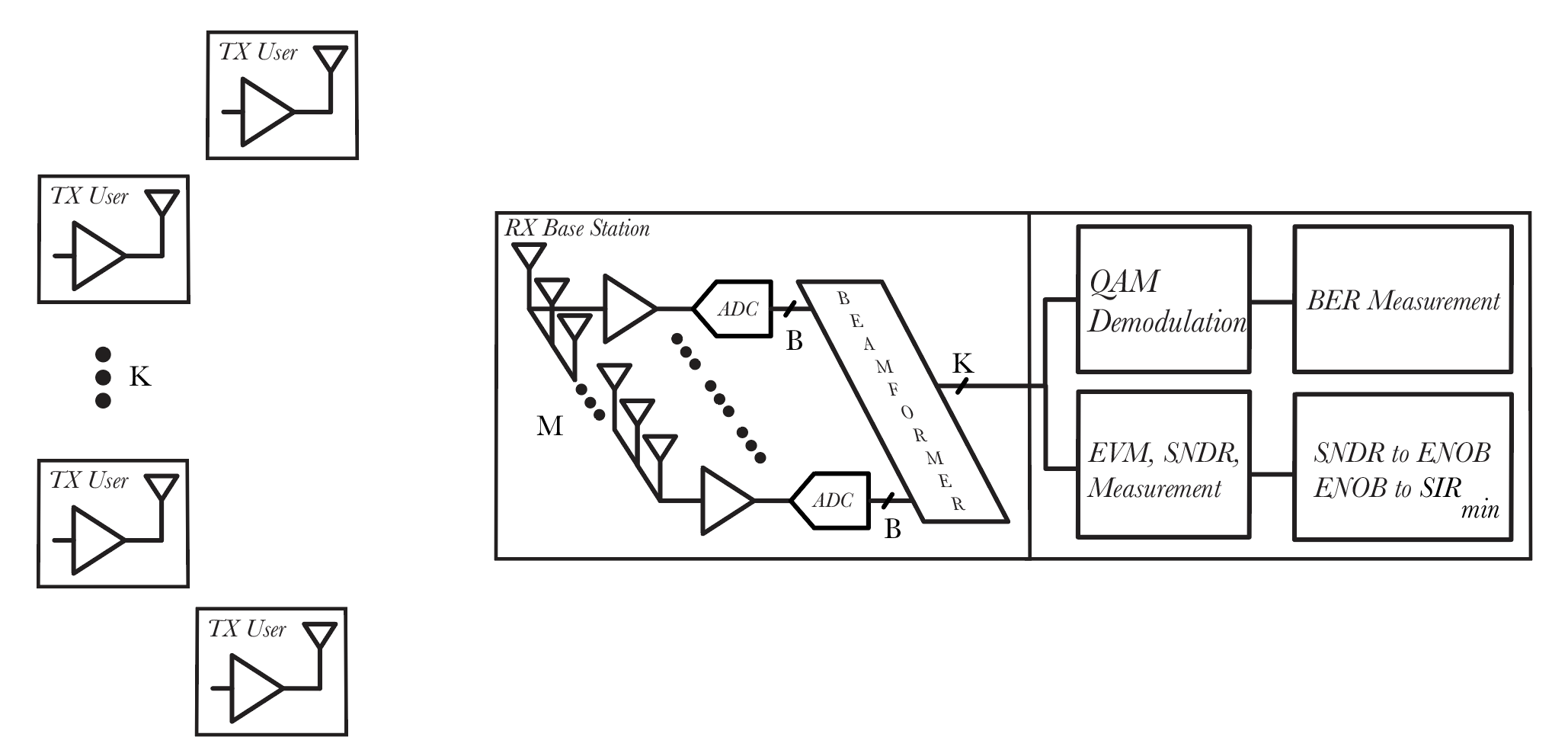}
    \caption{Overview of the Digital Beamforming Behavioral Model Implemented in MATLAB}
    \label{fig:intro_block_diagram}
\end{figure}

\subsection{Model Overview, Variable Names and Definitions}
Fig. \ref{fig:intro_block_diagram} shows an overview of the proposed behavioral model which simulates an end-to-end digital beamforming communication system. Each user independently transmits a random QAM signal over an ideal free-space line-of-sight channel. The M-element base station receives the signals in the far-field at each antenna element, quantizes each channel independently and performs beamforming in the digital domain. The beamforming operation, since fully digital, outputs a stream of data for each user.

Each of the output data streams is then independently processed by the measurement sub-block to determine BER, EVM and ENOB. These measurements can the be used to estimate ENOB, \(SIR_{min}\) and the array gain. 

Below are the names and definitions for the sweep variables, the boolean variables and the model outputs for plotting.

\subsubsection{Tunable Model Variables (Sweep Variables):}
\begin{itemize}
    \item \( \textbf{M} \): number of receive (base station) elements
    \item \( \textbf{K} \): number of transmitters (users or  interferer), each modeled as single elements. In a system where an interferer is present, there are K-1 users of interest and one interferer.
    \item \( \textbf{B}\): analog-to-digital converter (ADC) resolution
    \item \( \mathbf{SNR_{therm}} \): signal-to-thermal-noise ratio (in dB)  defined at the output of the LNA in figure \ref{fig:intro_block_diagram}
    \item \( \textbf{SIR} \): signal-to-interferer ratio (in dB). In a system with K users, there are K-1 users of interest transmitting at a nominal power level and one interfering user transmitting at a power level commensurate with the SIR.
    \item \( \textbf{Angle} \): Blocker or user angle sweep (in degrees). Depends on boolean variables  SweepUserAngle, SweepBlockerAngle.
\end{itemize}

\subsubsection{Boolean Model Variables:}
\begin{itemize}
    \item \( \textbf{LogScaleX, LogScaleY, LogScaleZ} \): determine the plot axes scale. If false, axes default to linear scale.
    \item \( \textbf{EnableBlocker}  \): if true, then one out of the K transmitting elements transmits at higher power depending on SIR. Must be enabled to sweep interferer angle.
     \item \( \textbf{SweepUserAngle, SweepBlockerAngle} \): determine whether the desired angle sweep will sweep the user angle spacing or the interferer angle. If both are selected, the model will sweep the interferer angle only. 
    \item \( \mathbf{ZF_{on}} \): if true, implements a zero-force beamforming. If false, implements conjugate beamforming. Note that the zero-forcing algorithm requires that $(M\geq K)$ and that users are sufficiently far apart. A rule of thumb is the user angular spacing must be greather than $\frac{\pi}{M}$ rad.
    \item \( \textbf{addThermNoise}  \): if false, does not add thermal noise at the receiver. 

\end{itemize}

\subsubsection{Model Outputs for Plotting}
\begin{itemize}
    \item \( \textbf{BER} \): the bit-error-rate and is defined as the ratio of the number of bits received incorrectly to the total number of bits transmitted during a specific period of time.
    \item \( \textbf{EVM} \): the error-vector magnitude and is a measure of the difference between the ideal and actual received signal constellation points in a communication system, expressed as a RMS percentage.
    \item \( \textbf{SNDR} \): the signal-to-noise-and-distortion ratio (in dB) and is a measure of the quality of a signal in the presence of both noise and distortion. 
    \item \( \textbf{ENOB} \): the effective number of bits that an ADC uses to quantize the signal of interest, rather than the noise and distortion present, and can be directly calculated from an SNDR measurement.
    \item \( \mathbf{SIR_{min}} \): the minimum signal-to-interferer ratio (in dB) that a particular beamforming system can handle. It is a heuristic that must be carefully chosen for a particular system. Above the minimum SIR, the strength of the interferer degrades the performance of the system below acceptable levels.
    \item \( \textbf{Array Gain} \): the ratio between the measured SNDR at the array output and the individual array channel. In an ideal phased array, the array gain is \(10 \cdot log_{10}(M)\)
\end{itemize}

\begin{figure}
    \centering
    \includegraphics[width=1\linewidth]{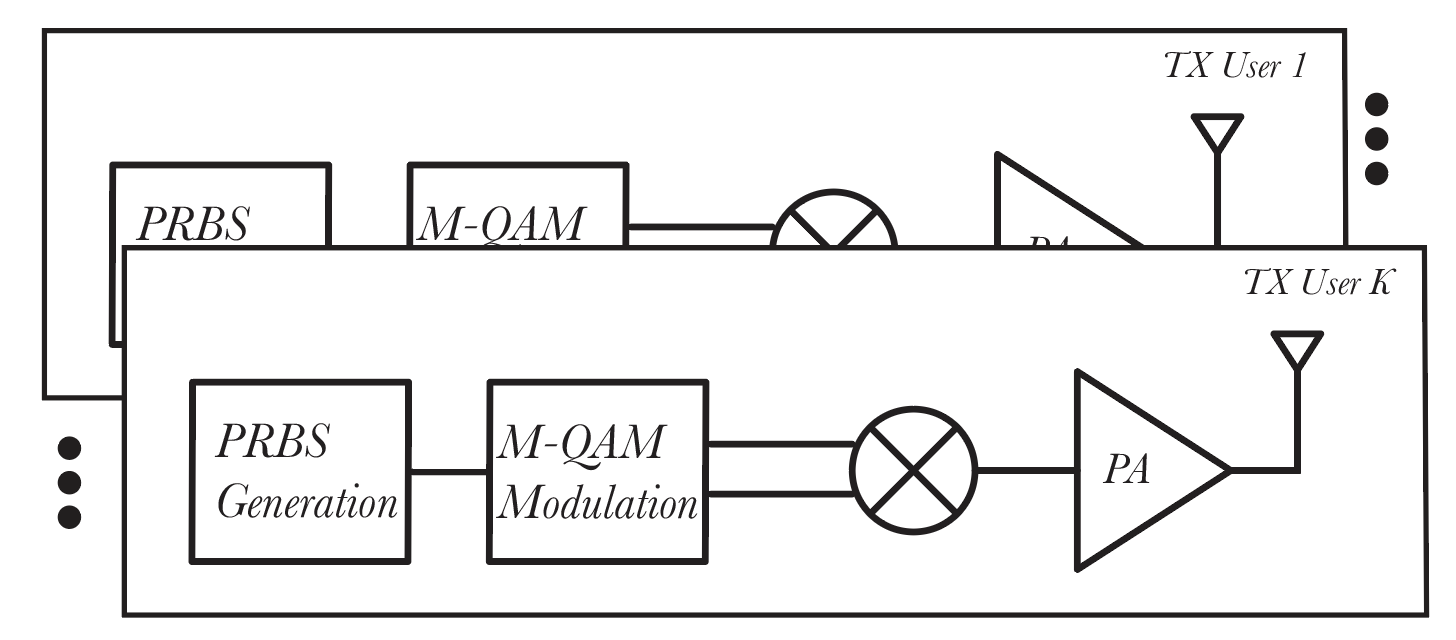}
    \caption{Transmitter Block Diagram}
    \label{fig:TX_block_diagram}
\end{figure}

\subsection{Modeling Transmitting User Elements}
Fig. \ref{fig:TX_block_diagram} shows a detailed block diagram of each transmitting user element. Firstly, a pseudo-random binary sequence (PRBS) is independently generated. Then, the bit sequence is modulated into 16-QAM data stream. Though depicted in the block diagram in Fig. \ref{fig:TX_block_diagram}, the mixer itself is omitted in the model since the data's complex envelope is being used. In other words, when operating on complex envelope data streams, the data streams themselves are already at baseband. Lastly, the signal is amplified by a linear power amplifier (non-linearities not modeled).

\begin{figure}
    \centering
    \includegraphics[width=1\linewidth]{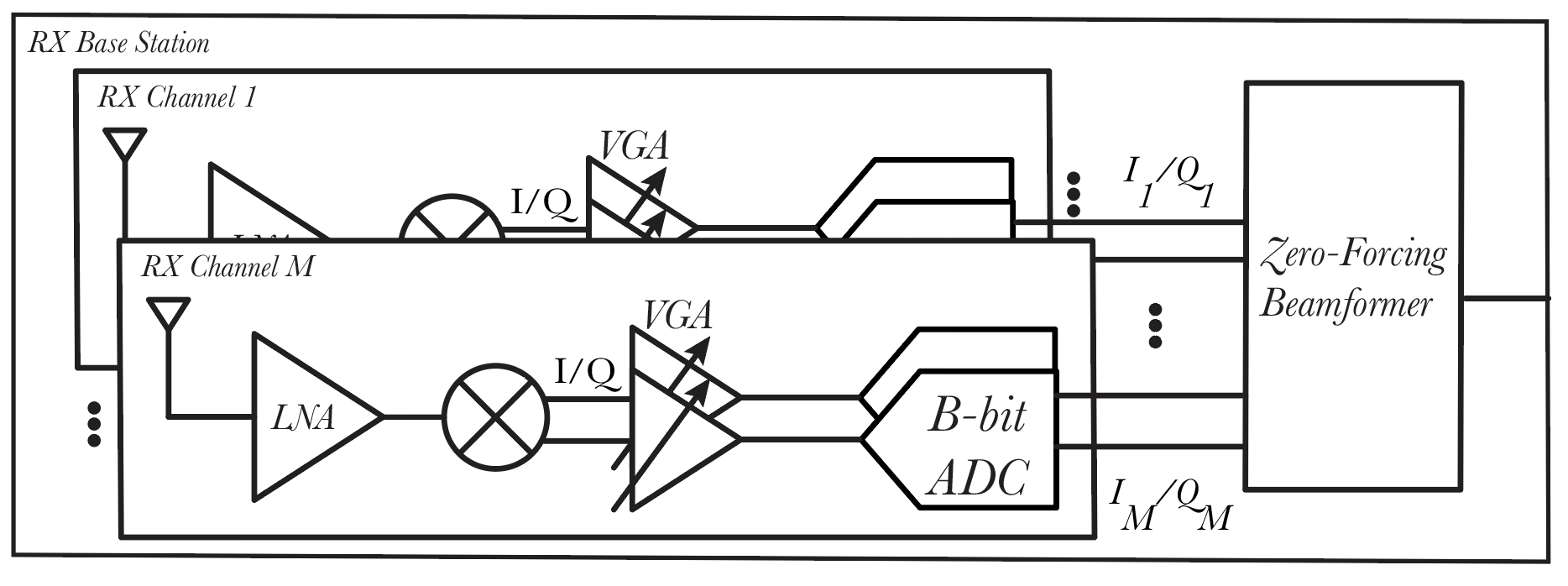}
    \caption{Receiver Block Diagram}
    \label{fig:RX_block_diagram}
\end{figure}

\subsection{Modeling Receiving Base Station}
Fig. \ref{fig:RX_block_diagram} shows a detailed block diagram of the M-element receiving base station modeled in this paper. Each receive channel consists of a low-noise amplifier, a variable gain amplifier, and a B-bit ADC. The low-noise amplifier is modeled as an additive white Gaussian noise (AWGN) source and a linear amplification. The variable gain amplifier is implemented by a normalization function that normalizes the average power of the received constellation to that of an ideal constellation. Lastly, the B-bit ADC is implemented in code as a mid-tread ADC. This is followed by the zero-force or conjugate beamforming operation that is further described in the next sub-section.

\begin{figure}
    \centering
    \includegraphics[width=0.9\linewidth]{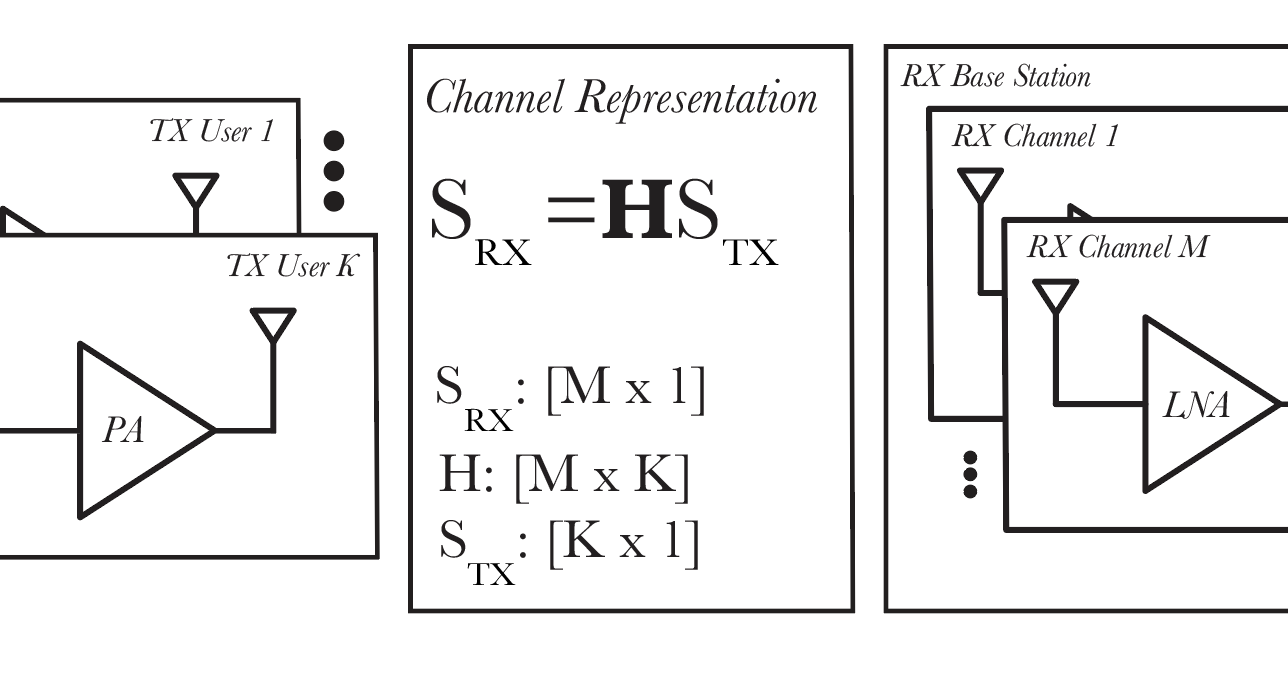}
    \caption{MIMO Channel Block Diagram}
    \label{fig:channel_block_diagram}
\end{figure}

\subsection{Modeling MIMO Channel and Beamforming Operation}
Fig. \ref{fig:channel_block_diagram} shows the MIMO channel diagram. The expected channel behavior is a progressive angle-dependent phase shift between elements with signal attenuation due to free-space path loss. As shown in Fig. \ref{fig:channel_block_diagram}, $S_{TX}$ and $S_{RX}$
are the signals at each antenna interface and are the inputs and outputs of the MIMO channel, respectively:

\begin{equation}
    S_{RX} = \mathbf{H}S_{TX}
\end{equation}

Ideal channel estimation is assumed, and the zero-force and conjugate beamforming matrices are computed as follows. Let $S_{RX}$ be the $M \times1$ vector of signals at each receive antenna and $R_{RX}$ be the $K \times 1$ vector of reconstructed desired signals:
\begin{equation}
    R_{RX} = \mathbf{G_{RX}}S_{RX}
\end{equation}

$G_{RX}$ is a matrix constructed from the channel estimate, $H$, depending on the objective of choice. In the case of conjugate beamforming:

\begin{equation}
    \mathbf{G_{RX,conj}} = \mathbf{H}^H
\end{equation}

In the case of zero-force beamforming: 
\begin{equation}
    \mathbf{G_{RX,ZF}} = (\mathbf{H}^H \mathbf{H})^{-1} \mathbf{H}^H
\end{equation}

We note that $\mathbf{H}^H$ represents the Hermitian, or conjugate, transpose of $\mathbf{H}$. The conjugate beamforming matrix, $\mathbf{G_{RX,conj}}$, maximizes the beamformed signal energy while the zero-force beamforming matrix, $\mathbf{G_{RX,ZF}}$, prioritizes nulling out up to $M-1$ unwanted interference sources but may trade off signal strength to do so \cite{puglielli}.

A future area of improvement for this model is including a frequency-dependent fading channel model, building on the fairly simple beamforming channel model used here.

\subsection{Measurement Blocks}
\subsubsection{Measuring SNDR}

\begin{figure}
    \centering
    \includegraphics[width=1\linewidth]{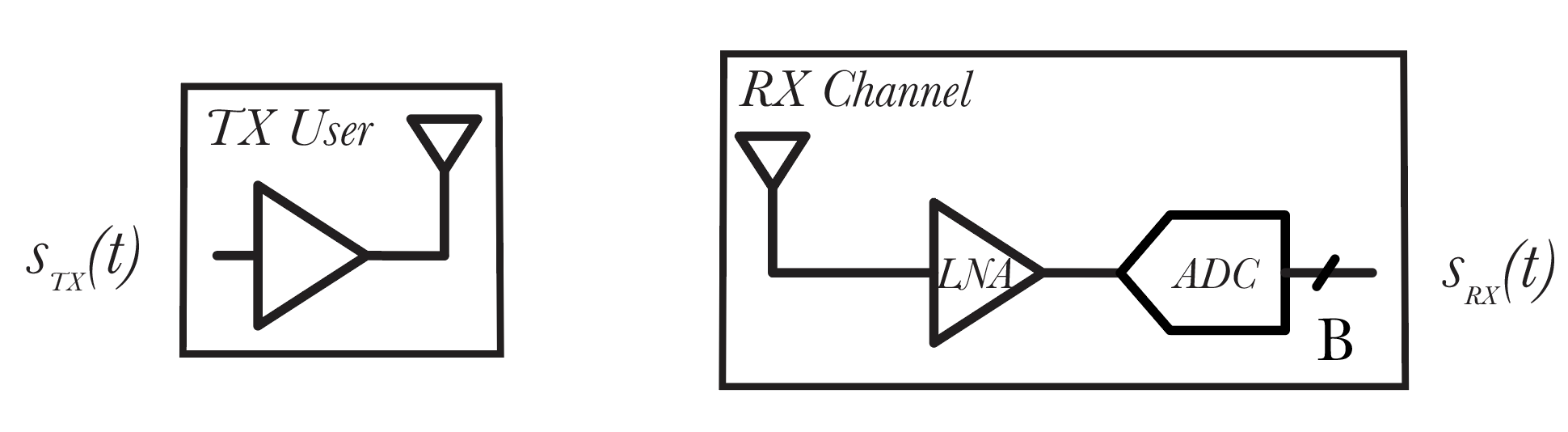}
    \caption{Single-User, Single-Element Uplink}
    \label{fig:single_channel}
\end{figure}

The SNDR of the output signal can be "measured" as follows. Take as an example the uplink depicted in Fig. \ref{fig:single_channel}. $s_{TX}(t)$ and $s_{RX}(t)$ are the transmitted and received QAM-modulated signals of interest, respectively. Note that $s_{TX}(t)$ is an ideal noiseless signal and $s_{RX}(t)$ has been degraded by noise and distortion from input to output. In other words, $s_{TX}(t)$ is the "ground truth" signal of interest. In order to determine the SNDR of $s_{RX}(t)$, we may compute $C$, the cross-correlation coefficient between $s_{TX}(t)$ and $s_{RX}(t)$. In this case, C is the maximum cross-correlation and does not necessarily occur at zero lag due to potential phase delays from TX to RX.

\begin{equation}
    C = crossCorr(s_{RX}(t), s_{TX}(t))
\end{equation}

Let us define the SNDR as follows, where $P_{RX,S}$ and $P_{RX,S}$ are the signal and noise/distortion power of the received signal, respectively: 

\begin{equation}
    SNDR = 10log_{10}(P_{RX,S}/P_{RX,N+D})
\end{equation}

We note that the total power of $s_{RX}(t)$ is the sum of the signal and noise/distortion and is proportional to the variance  $s_{RX}(t)$:
\begin{equation}
    (P_{RX,S} + P_{RX,N+D}) \propto var(s_{RX}(t))
\end{equation}

Additionally, since $P_{RX,S}$ represents the strength of the originally transmitted signal that is still carried by $s_{RX}(t)$:
\begin{equation}
    P_{RX,S} \propto |C|^2 var(s_{RX}(t))
\end{equation}
Lastly, we note that the noise/distortion power $P_{RX,N+D}$ can be approximated. $|C|s_{TX}(t)$ represents the part of $s_{RX}(t)$ that fully correlates with $s_{TX}(t)$. In other words $|C|s_{TX}(t)$ is the signal of interest. When subtracted from $s_{RX}(t)$, we may compute $P_{RX,N+D}$: 
\begin{equation}
    P_{RX,N+D} \propto var(s_{RX}(t) - |C|s_{TX}(t))
\end{equation}

Thus,

\begin{equation}
    SNDR = 10log_{10}(\frac{|C|^2 var(s_{RX}(t))}{var(s_{RX}(t) - |C|s_{TX}(t))})
\end{equation}

In summary, the beamformed signal will contain noise and distortion. Using the cross-correlation function, we can estimate the strength of the original signal power present in the beamformed signal. Then, the difference between the total power and the signal power is the noise and distortion power. The ratio of these two yields the SNDR. Note that the maximum cross-correlation (rather than the cross-correlation at zero lag) must be calculated since the beamformed signal will often be out of phase with respect to the transmitted signal.

\subsubsection{Computing ENOB for QAM Signals}

For a sinusoidal signal, the following expression is generally used to compute ENOB:
\begin{equation}
    ENOB = (SNDR - 1.76)/6.02
\end{equation}

However, this expression is not accurate for QAM-modulated signals or signals that experience matched filtering. Rather, the peak-to-average-power ratio (PAPR) affects the ENOB and must be accounted for as is discussed in \cite{reynaert}. 

In general:
\begin{equation}
    ENOB = (SNDR - C)/6.02
\end{equation}

Where C depends on the PAPR of the constellation of interest and varies with modulation order and with matched filtering parameters. For example, for 16-QAM without matched filtering:
\begin{equation}
    ENOB = (SNDR - 4.36)/6.02
\end{equation}

\subsubsection{A New Heuristic: $\mathbf{SIR_{min}}$}
In order to determine the performance of the system in the presence of interference, we introduce a new heuristic - $SIR_{min}$. As its name suggests, $SIR_{min}$ represents the minimum tolerable signal-to-interferer ratio. In other words, for an SIR smaller than $SIR_{min}$, the interferer strength degrades system performance beyond what would be considered acceptable.

In order to properly define $SIR_{min}$, a modulation scheme must be assumed. For all modeling results shown in this paper, 16-QAM modulation is assumed. It has been observed that, under general conditions, 16-QAM modulation requires a minimum ADC resolution of two bits. Intuitively, this can be justified since, in order to represent sixteen unique QAM constellation points, two bit resolution is needed in each of the I and Q paths. 

As a result, we define $SIR_{min}$ as the minimum SIR above which the receiver ENOB is greater than two. In other words, we consider any SNDR degradation beyond an ENOB of two to be prohibitively detrimental to system performance. 

\subsubsection{Effective Array Gain}
The effective array gain is computed by subtracting the channel SNR (before the ADC) from the measured output SNDR as follows:
\begin{equation}
    AG = SNDR_{out} - SNDR_{in}
\end{equation}

Note that $SNDR_{in}$ is taken at the LNA output, before the signal is digitized. This is done deliberately to include the effects of ADC quantization on the effective array gain.

\subsection{Equation-based Model of SNDR Degradation due to Interference}

\begin{figure}
    \centering
    \includegraphics[width=1\linewidth]{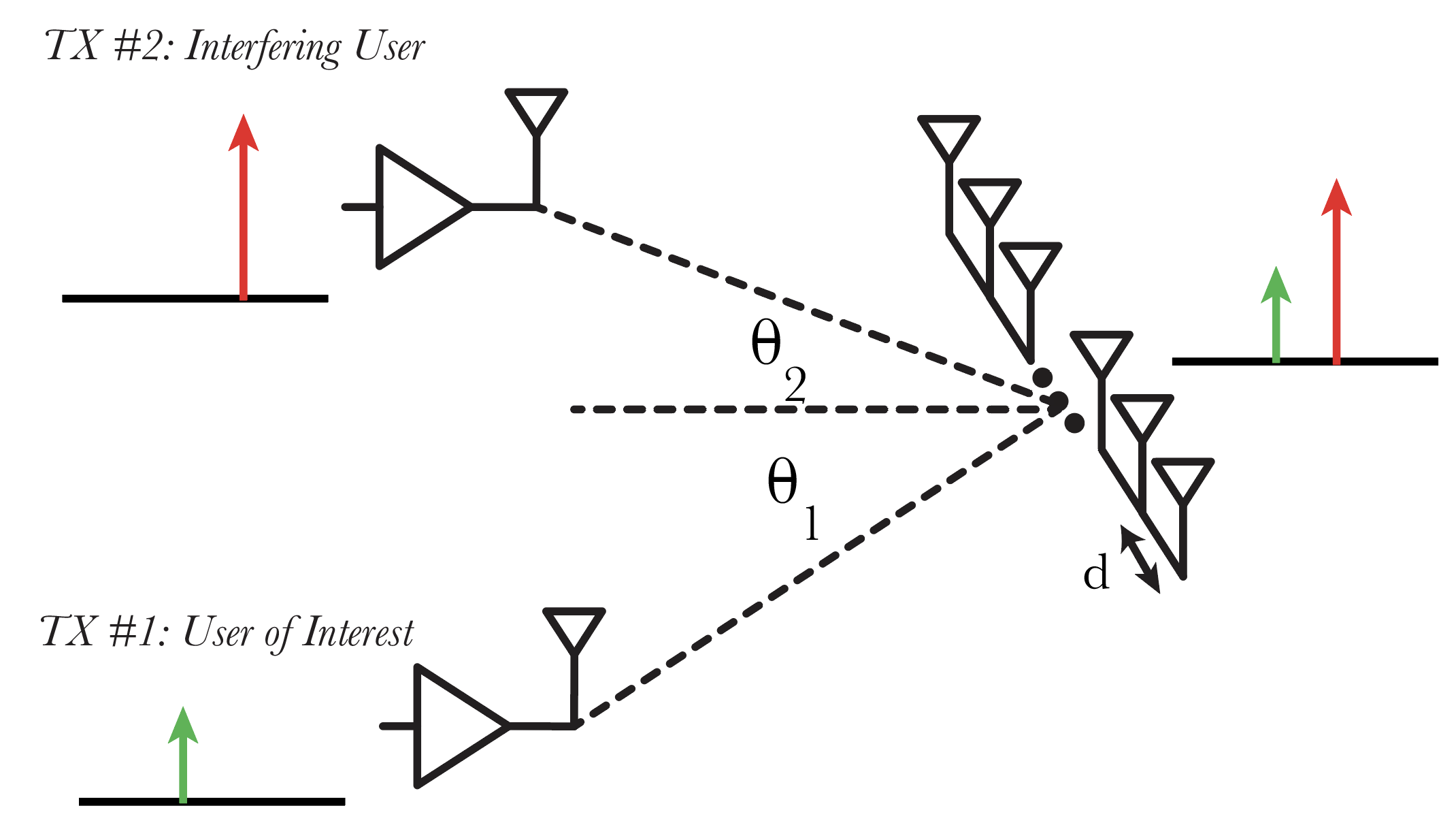}
    \caption{Single-Element Uplink with a User of Interest and an Interfering User}
    \label{fig:twoWavefronts}
\end{figure}

Before diving deeply into the results of the behavioral model, an equation-based model of the effects of an interferer on a single-channel receiver is presented. The equation-based model presented below forms the basis for what we may expect to observe in the behavioral model.

Consider the receiver shown in Fig. \ref{fig:twoWavefronts} receiving two wavefronts - one from a user of interest and the other from an interfering signal. Let's assume that the receiver produces no thermal noise so  that the output SNDR will be dominated by either 1) the quantization noise of the downstream B-bit ADC (not shown in Fig. \ref{fig:twoWavefronts}) or 2) the interferer power itself:

\begin{equation}
    SNDR = \frac{P_{sig}}{P_{q,noise}+P_{interf}} = \frac{1}{\frac{1}{SQNR}+\frac{1}{SIR}} = (SQNR||SIR)
    \label{eqn:SNDRlinear}
\end{equation}

Note that in Eqn. \ref{eqn:SNDRlinear} SNDR, SQNR and SIR are written in linear scale. We can further rewrite SQNR, now in dB scale, as follows:

\begin{equation}
    SQNR_{dB} = 6.02B -1.76 - 10log_{10}(\frac{P_{FS}}{P_{sig}})
    \label{eqn:SQNRwithFS}
\end{equation}

where $P_{FS}$ is the full-scale power of the ADC and $P_{sig}$ is the power level of the signal of interest. In the absence of interference, we assume that the ADC full-scale can be entirely  allocated to the signal of interest. In this case, when $P_{FS} = P_{sig}$, the last term in Eqn.  \ref{eqn:SQNRwithFS} is zero: 

\begin{equation}
    SQNR_{dB} = SQNR_{nom, dB} = 6.02B -1.76 
\end{equation}

However, in the presence of an interferer, the full-scale must be shared by the signal of interest and the interferer. If we assume that the interfering signal is uncorrelated with respect to the signal of interest, the signals combine in power. In this scenario, $P_{FS} = P_{sig} + P_{interf}$, where $P_{interf}$ is the interferer power level. Rewriting Eqn. \ref{eqn:SQNRwithFS},

\begin{equation}
    SQNR_{dB}= 6.02B -1.76 - 10log_{10}(\frac{P_{sig}+P_{interf}}{P_{sig}})
\end{equation}

\begin{equation}
    SQNR_{dB} = SQNR_{nom,dB} - 10log_{10}(\frac{P_{sig}+P_{interf}}{P_{sig}})
\end{equation}

\begin{equation}
    SQNR_{dB} = SQNR_{nom} - 10log_{10}(1+\frac{1}{SIR})
    \label{eqn:SIRdegradesSQNR}
\end{equation}

\begin{figure}
    \centering
    \includegraphics[width=0.9\linewidth]{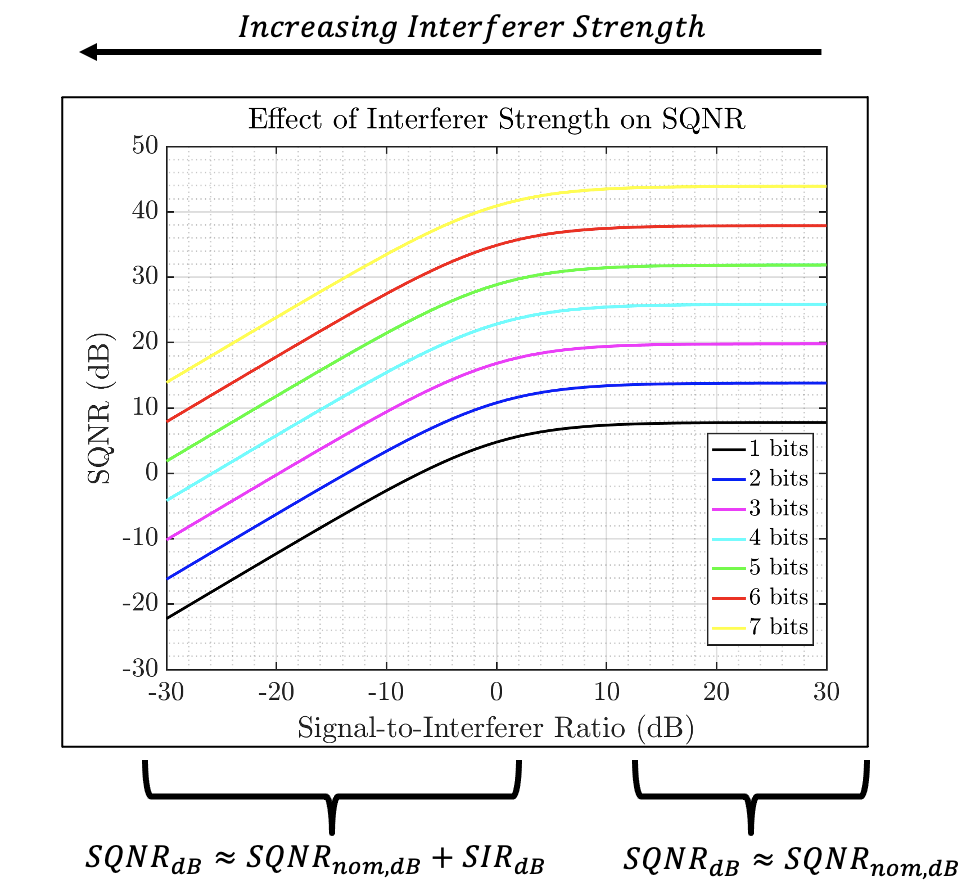}
    \caption{Plot of Eqn. \ref{eqn:SIRdegradesSQNR} Showing SQNR Degradation in the Presence of Strong Interferers}
    \label{fig:SQNRvsSIRIdeal}
\end{figure}

As shown in Fig. \ref{fig:SQNRvsSIRIdeal}, and described by Eqn. \ref{eqn:SIRdegradesSQNR}, the effective quantization noise added by the ADC in a receiver affected by interference is degraded by the interferer power itself. That is, even if the interferer signal were nulled by the beamformer after digitization, we would expect that the output SNDR would still degrade since the effective quantization noise is degraded by the presence of the interferer. Intuitively, this can be understood by observing that the ADC full-scale must be shared by both the signal of interest and the interferer, effectively degrading the dynamic range allocated by the ADC to the signal of interest. Taking note of this observation is instructive to analyze the results of the behavioral model presented below.

\section{MATLAB Behavioral Model Results}
\label{section:ChDBFResults}
\label{ChDBFModelResults}

As has been previously described, the behavioral model presented in this paper has several tunable and sweepable variables and allows various outputs to be plotted. This section presents several key results that describe how various system parameters affect system performance in the presence of a strong interferer. 

Firstly, the system is modeled without thermal noise to independently model the effects of interference on quantization noise. Then, the model is utilized to determine a system's minimum tolerable SIR and optimum ADC resolution. Lastly, some additional results on one-bit ADC topologies user count effects and thermal noise effects are presented.  

\subsubsection{Quantization Noise Degradation in the Presence of Strong Interferers}

Digital beamforming systems with low-resolution ADCs are of critical importance to enable low-power MIMO. Therefore, it is instructive to analyze the effects of systems where the quantization noise is significant.

\begin{figure}
    \centering
    \includegraphics[width=1\linewidth]{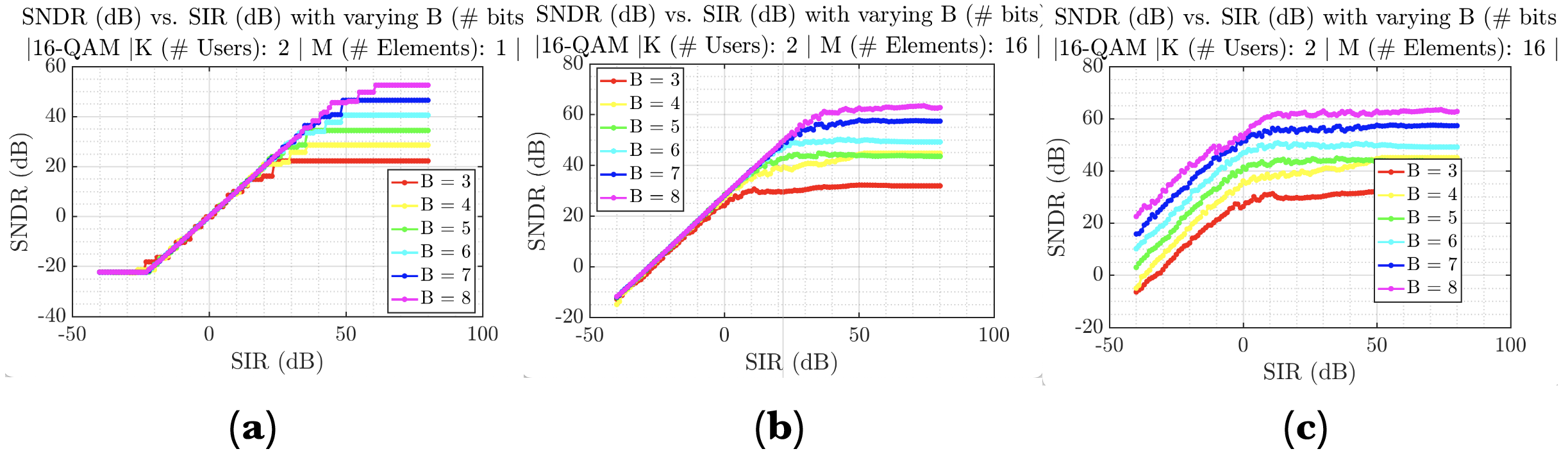}
    \caption{SNDR vs. SIR for a) M = 1, b) M = 16 with conjugate beamforming c) M = 16 with zero-force beamforming}
    \label{fig:SNDRvsSIR_1_16_CBF_ZF}
\end{figure}

Fig. \ref{fig:SNDRvsSIR_1_16_CBF_ZF} plots the measured SNDR of a 16-QAM signal at the receiver's beamformed output in the presence of an interferer and in the absence of thermal noise  for a) single-element receiver, b) sixteen-element conjugate beamforming receiver and c) sixteen-element zero-force beamforming receiver. The SNDR is plotted against SIR for various values of B, the ADC resolution.

The behavior described by Eqns. \ref{eqn:SNDRlinear} and \ref{eqn:SIRdegradesSQNR} can be observed in Fig. \ref{fig:SNDRvsSIR_1_16_CBF_ZF}a) and \ref{fig:SNDRvsSIR_1_16_CBF_ZF}b). In Fig. \ref{fig:SNDRvsSIR_1_16_CBF_ZF}a) when SIR is below zero, the interferer power is high and the SNDR is largely dominated by the SIR. When the SIR is greater than zero, the quantization noise of the ADC dominates the performance. The same behavior can be observed in Fig. \ref{fig:SNDRvsSIR_1_16_CBF_ZF}b) because the interferer appears at a sidelobe of the sixteen-element beampattern. 

However, as shown in Fig. \ref{fig:SNDRvsSIR_1_16_CBF_ZF}c), it is possible to partially mitigate the degradation due to the interferer by using a zero-force beamformer. The zero-force operation nulls the interferer signal so we expect that the SNDR at the output is solely dependent on the SQNR. As described by Eqn. \ref{eqn:SIRdegradesSQNR}, however, the SQNR itself is still degraded by the interferer strength.

Indeed, the behavioral model of the digital beamforming system can be utilized to validate the equation-based model discussed previously. In conjunction, the hypothesis described by equations Eqns. \ref{eqn:SNDRlinear} and \ref{eqn:SIRdegradesSQNR} has been validated by the behavioral model, enabling a deeper understanding of the effects of interference on beamformer performance.

\subsubsection{Adding Thermal Noise}

\begin{figure}
    \centering
    \includegraphics[width=1\linewidth]{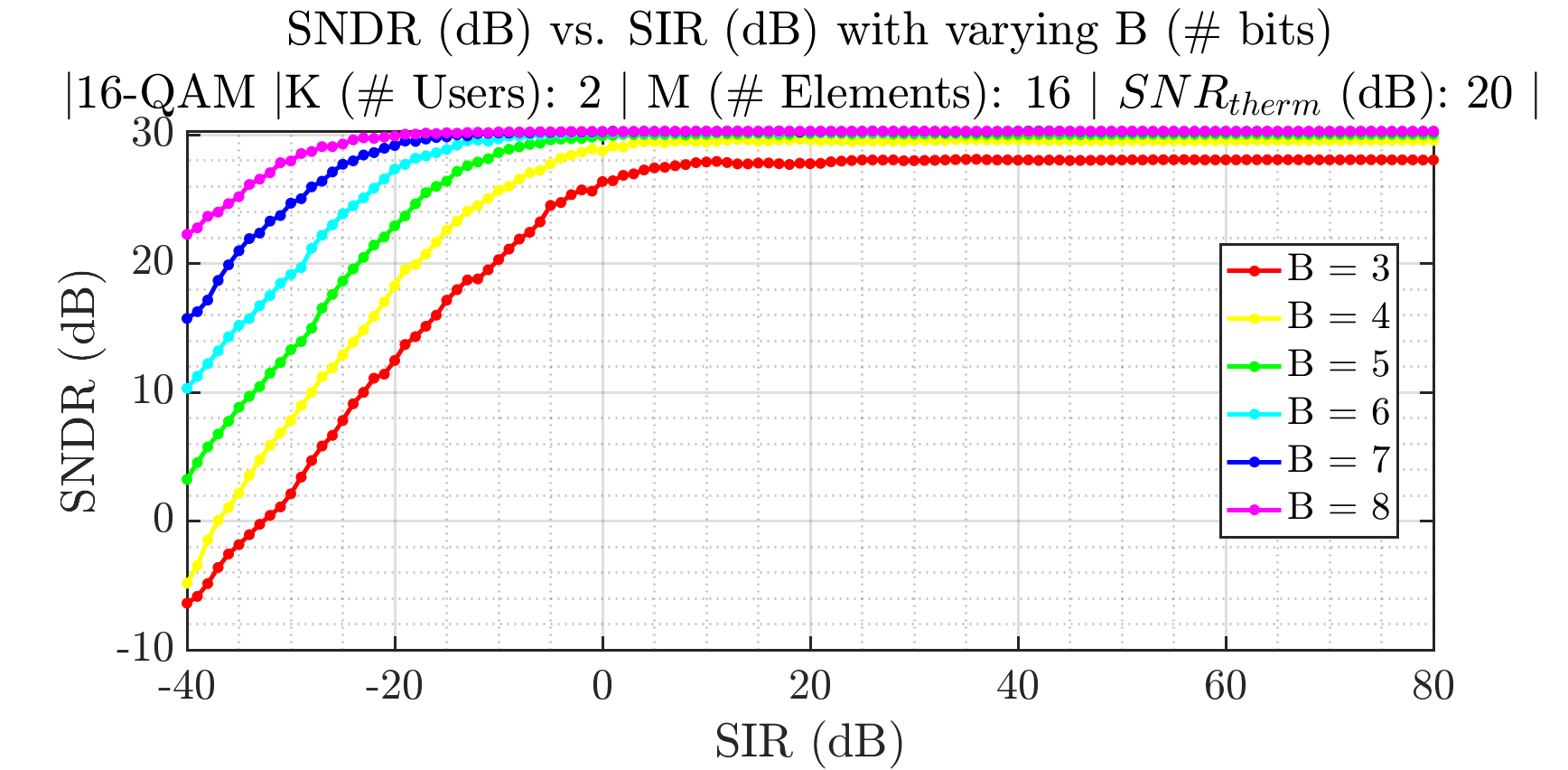}
    \caption{SNDR vs. SIR with M = 16, $SNR_{therm}$ = 20 dB, Zero-Force Beamforming}
    \label{fig:SNDRvsSIR_16ZF_Therm}
\end{figure}

Fig. \ref{fig:SNDRvsSIR_16ZF_Therm} shows the same zero-force beamformer described by Fig \ref{fig:SNDRvsSIR_1_16_CBF_ZF}c) but with thermal noise enabled. It can be seen that in the weak-interferer regime (SIR $>$ 0), the SNDR is dominated by thermal or quantization noise. For a low-resolution ADC, the quantization noise dominates and for a sufficiently high-resolution ADC the thermal noise will dominate. In Fig. \ref{fig:SNDRvsSIR_16ZF_Therm}, the thermal noise level is approximately 32 dB due to the $10log_{10}(M)$ SNR boost of the array.  The key observation, however, is that in the presence of a strong interferer (SIR $<$ 0), the degraded quantization noise level is much higher than the thermal noise level and the SQNR dominates the overall SNDR. That is to say, in the presence of a strong interferer, the receiver is likely to be largely quantization-noise limited at the edge of viable operation.

\subsection{Determining Minimum SIR}

\begin{figure}
    \centering
    \includegraphics[width=0.9\linewidth]{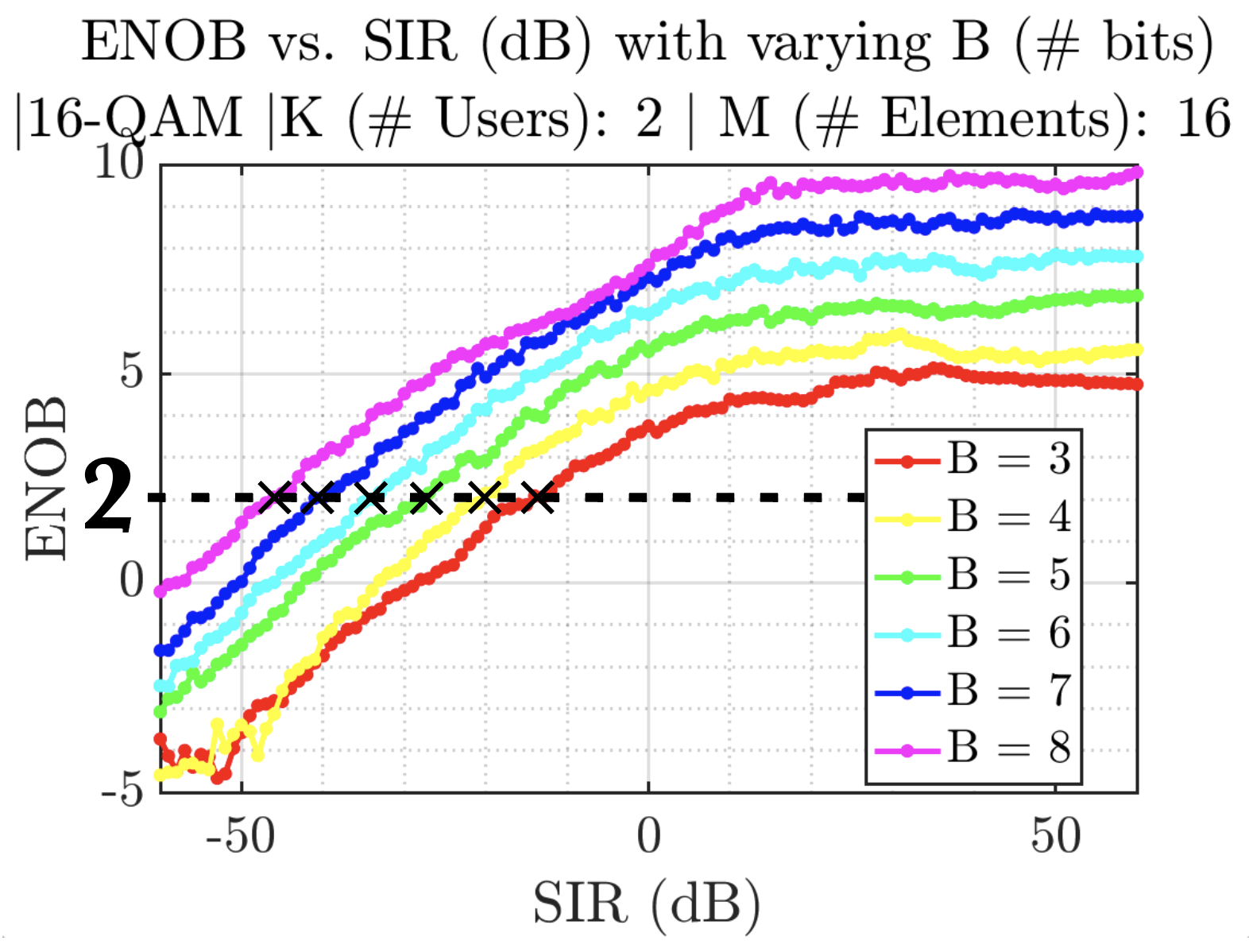}
    \caption{SNDR vs. SIR with M = 16, $SNR_{therm}$ = 20 dB, Zero-Force Beamforming. For each value of B, the intersection point of the ENOB vs. SIR curve with the horizontal line at ENOB = 2 denotes the $SIR_{min}$.}
    \label{fig:ENOBvsSIR_ENOB2}
\end{figure}

A primary motivation for developing this behavioral model was to understand the maximum tolerable interferer level, equivalently the minimum SIR, the system of interest could handle. As previously described, for the 16-QAM results shown in this paper the minimum tolerable SIR is defined as the SIR at which the ENOB degrades to two bits. Fig. \ref{fig:ENOBvsSIR_ENOB2} visualizes this definition for a sixteen-element zero-force beamformer and highlights that the minimum tolerable SIR varies with ADC resolution. For now, we neglect thermal noise as we have previously made the general observation that the degraded quantization noise level tends to dominate the overall performance when the interferer is strong.

By defining $SIR_{min}$ as we have, we can use the curves shown in Fig. \ref{fig:ENOBvsSIR_ENOB2} to computationally determine $SIR_{min}$ by extrapolating the intersection points between the ENOB vs. SIR curve and the ENOB = 2 horizontal line.

\begin{figure}
    \centering
    \includegraphics[width=0.9\linewidth]{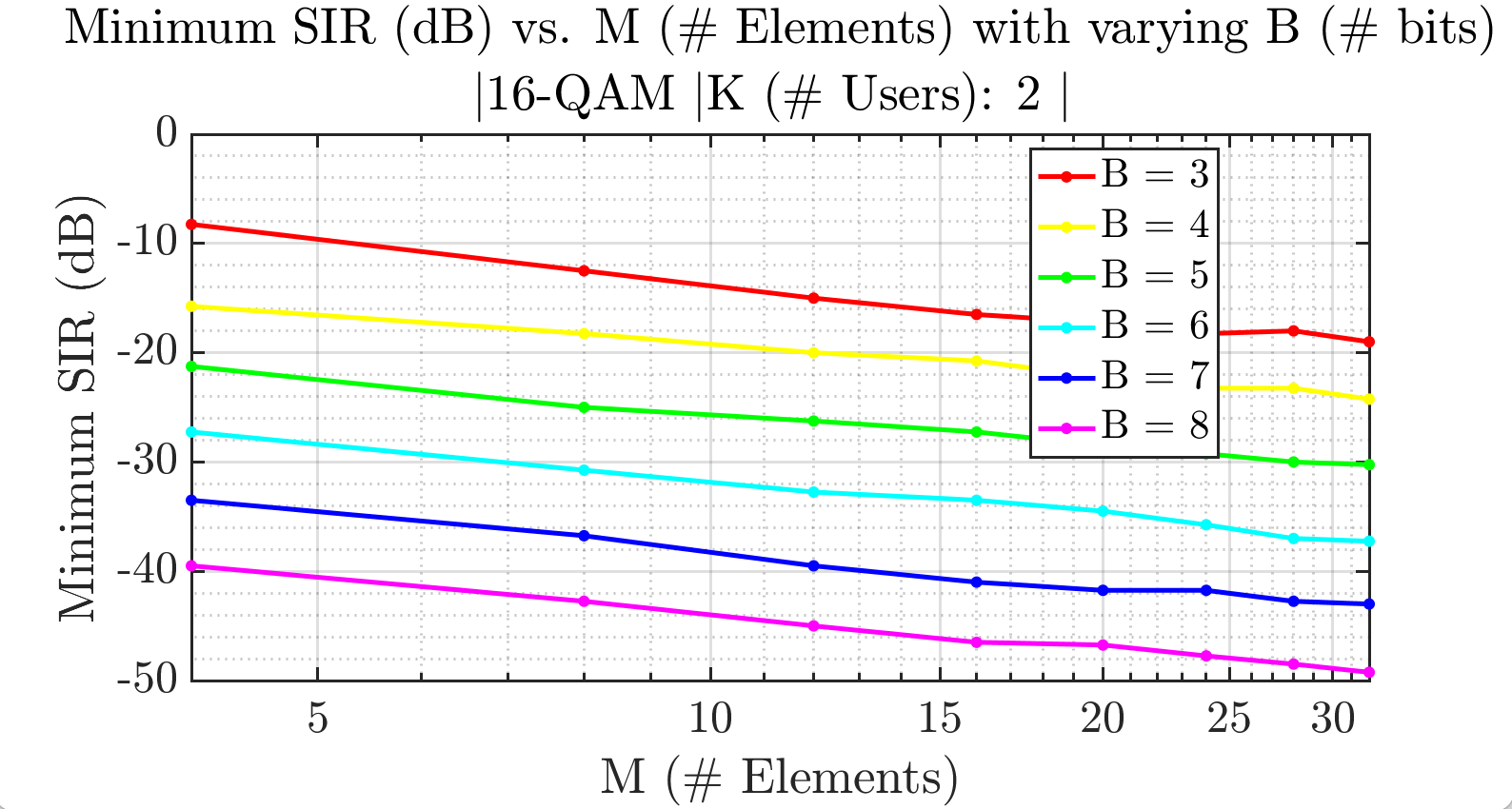}
    \caption{Minimum SIR vs. M for varying B. We observe trends of 6 dB improvement per bit and 3 dB improvement as the array size doubles.}
    \label{fig:SIRminvsMK}
\end{figure}

Fig \ref{fig:SIRminvsMK} plots our defined $SIR_{min}$ when varying the number of antenna elements and the ADC resolution in the absence of thermal noise. It can be observed that the minimum SIR improves by 6 dB per additional bit of ADC resolution and improves by 3 dB as the number of antennas doubles. Intuitively, this can be justified as follows - generally, the minimum SIR occurs when the SNDR and ENOB are primarily dominated by the degraded SQNR described by Eqn.  \ref{eqn:SIRdegradesSQNR}. As a result, we'd expect that the minimum SIR improves as 6 dB per bit since the SQNR increases by 6 dB per bit. Similarly, the SQNR generally improves by 3 dB when the array size doubles. However, the SQNR does not always improve as the array scales. This is because the quantization noise is not necessarily uncorrelated across channels in the way that thermal noise is. This is the case, for example when the receiver is steered toward broadside -  the constellations receiver at each element are identical (not rotated) and, thus, the quantization noise is fully correlated across channels.

\subsection{Determining Minimum ADC Resolution}
A secondary purpose for developing this digital beamforming model was to determine the minimum ADC resolution required for a digital beamforming system to modulate 16-QAM data reliably. In general, our findings indicate that for modestly sized arrays an ADC resolution of 3-5 bits is sufficient. However, recent work has suggested that one-bit ADCs may be feasible in such arrays. A case study on digital beamformers with one-bit ADCs is also presented.
\subsubsection{Low-resolution ADCs for Digital Beamforming}

\begin{figure}
    \centering
    \includegraphics[width=0.9\linewidth]{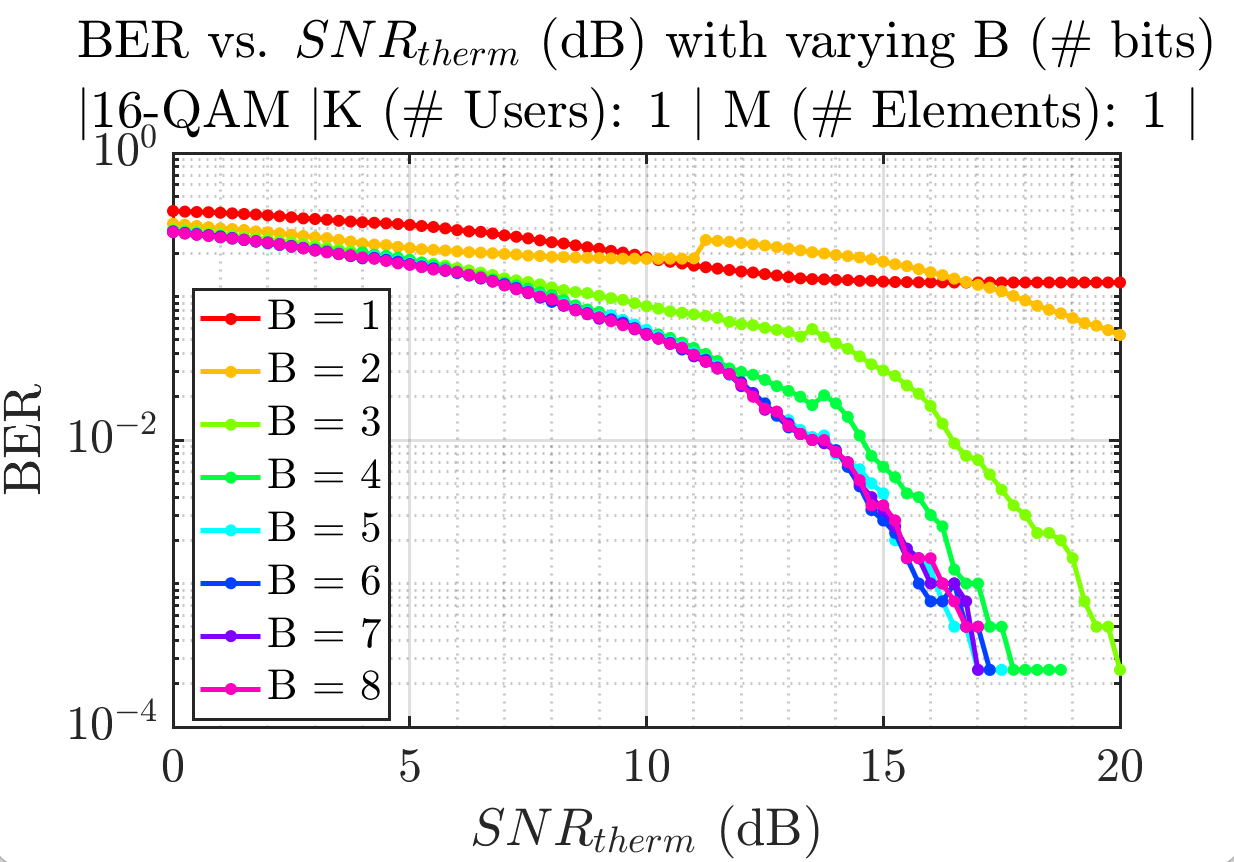}
    \caption{BER vs. $SNR_{therm}$ for varying B, M = 1}
    \label{fig:ChModeling2_BERvsSNRvsB_M1}
\end{figure}

\begin{figure}
    \centering
    \includegraphics[width=0.9\linewidth]{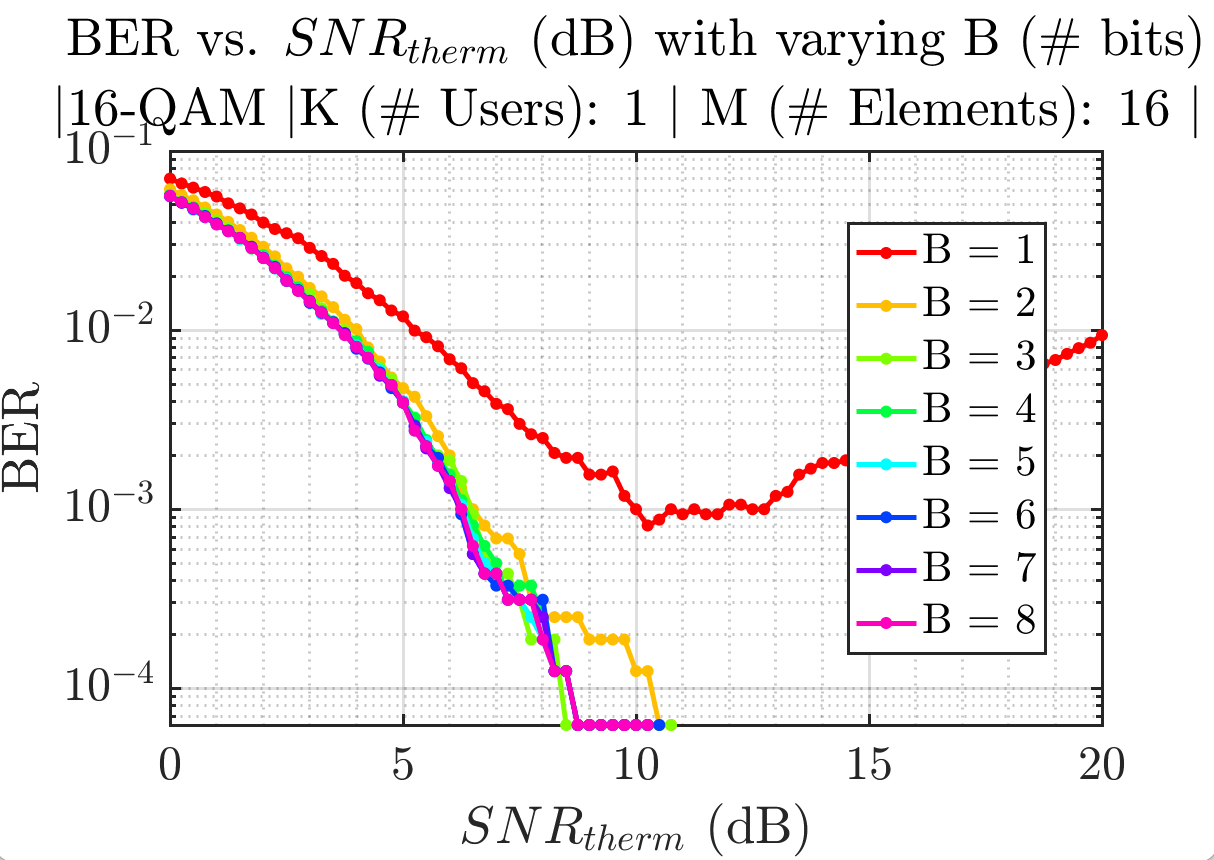}
    \caption{BER vs. $SNR_{therm}$ for varying B, M = 16}
    \label{fig:ChModeling2_BERvsSNRvsB_M16}
\end{figure}

One of the key benefits of digital beamforming architectures is that they generally require lower resolution ADCs than analog or hybrid beamforming counterparts. Fig. \ref{fig:ChModeling2_BERvsSNRvsB_M1} shows that indeed, a four-bit ADC is sufficient for 16-QAM modulation for a single-element receiver. As the array size grows, the minimum ADC resolution can further decrease. Fig. \ref{fig:ChModeling2_BERvsSNRvsB_M16} shows that for a sixteen-element receiver, ADC resolution as low as two bits may be viable.

\begin{figure}
    \centering
    \includegraphics[width=0.9\linewidth]{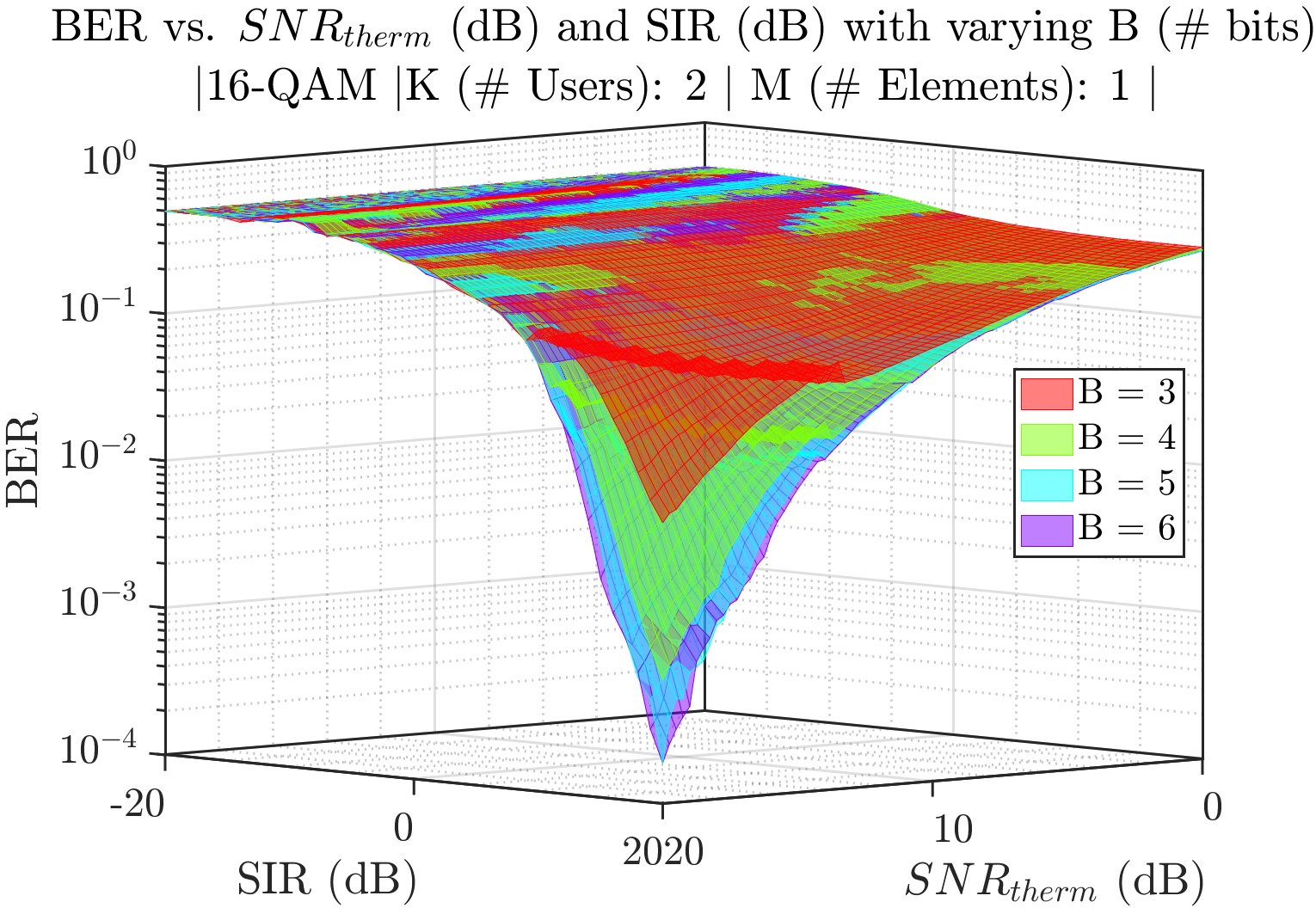}
    \caption{BER vs. $SNR_{therm}$ and SIR for varying B, M = 1}
    \label{fig:ChModeling2_BERvsSNRvsSIRvsB_M1}
\end{figure}

\begin{figure}
    \centering
    \includegraphics[width=1\linewidth]{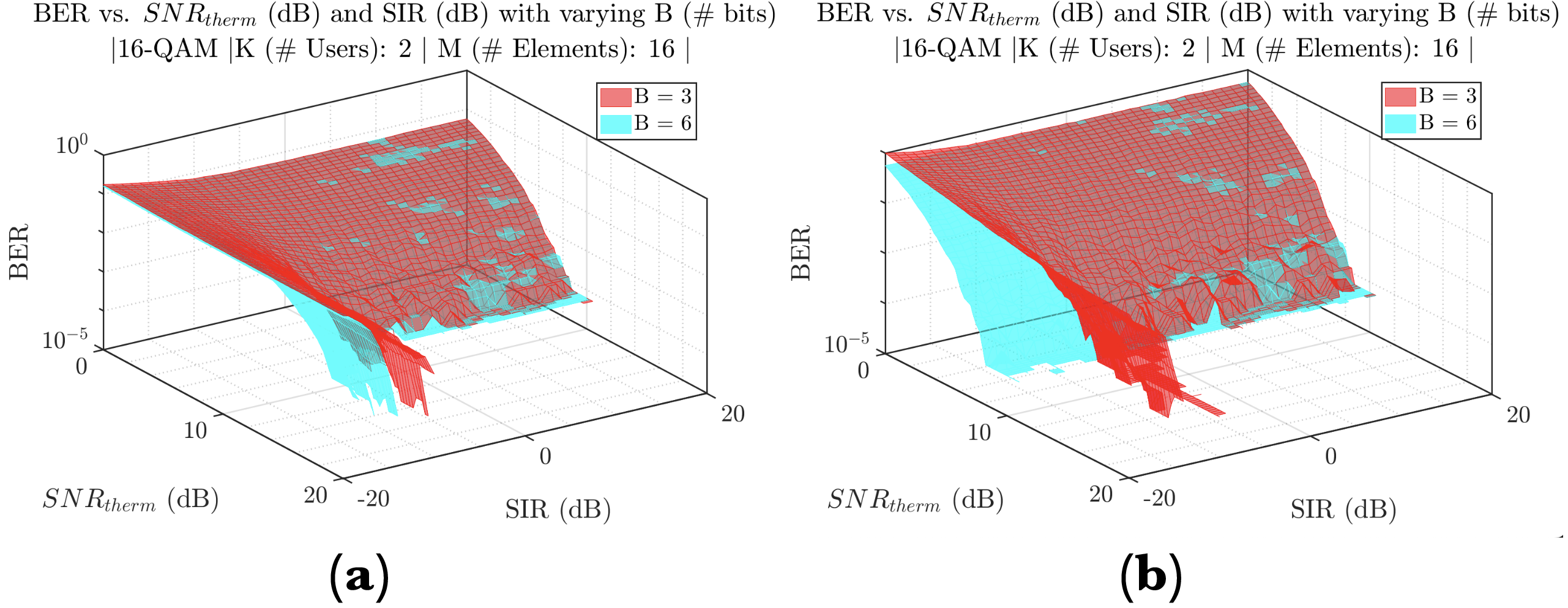}
    \caption{BER vs. $SNR_{therm}$ and SIR for varying B, M = 16, a) Conjugate Beamforming b) Zero-force Beamforming}
    \label{fig:ChModeling2_BERvsSNRvsSIRvsB_M16_CBF_ZFBF}
\end{figure}

However, in the presence of an interferer this is not necessarily the case. Fig. \ref{fig:ChModeling2_BERvsSNRvsSIRvsB_M1} shows that for a single-element receiver, the performance degrades heavily in the presence of a strong interferer regardless of ADC resolution. Extending this to an array, Fig. \ref{fig:ChModeling2_BERvsSNRvsSIRvsB_M16_CBF_ZFBF} shows a similar plot for a) sixteen-element conjugate beamformer and b) sixteen-element zero-force beamformer. In general, it can be observed that conjugate beamformers are very susceptible to interference, similar to the single-element case. On the other hand, a zero-force beamformer that is able to spatially null the interferer can operate with ADC resolution as low as three bits. However, it is important to note that the performance in such a case heavily depends on the strength of the interferer and the size of the array. In the case presented in Fig. \ref{fig:ChModeling2_BERvsSNRvsSIRvsB_M16_CBF_ZFBF}b), the sixteen-element array with three-bit ADCs can handle SIR as low as -10 dB.

\subsubsection{Case Study: Digital Beamforming with One-bit ADCs}
One-bit ADCs for digital beamforming have been extensively studied. One key result that has been previously published is that one-bit ADCs are capable of maintaining a communication link with complex modulation schemes, such as 16-QAM \cite{studer}. At first glance, this result is counter-intuitive; how could a 16-point constellation be demodulated by two one-bit ADCs that can only quantize to four constellation points?

\begin{figure}
    \centering
    \includegraphics[width=0.9\linewidth]{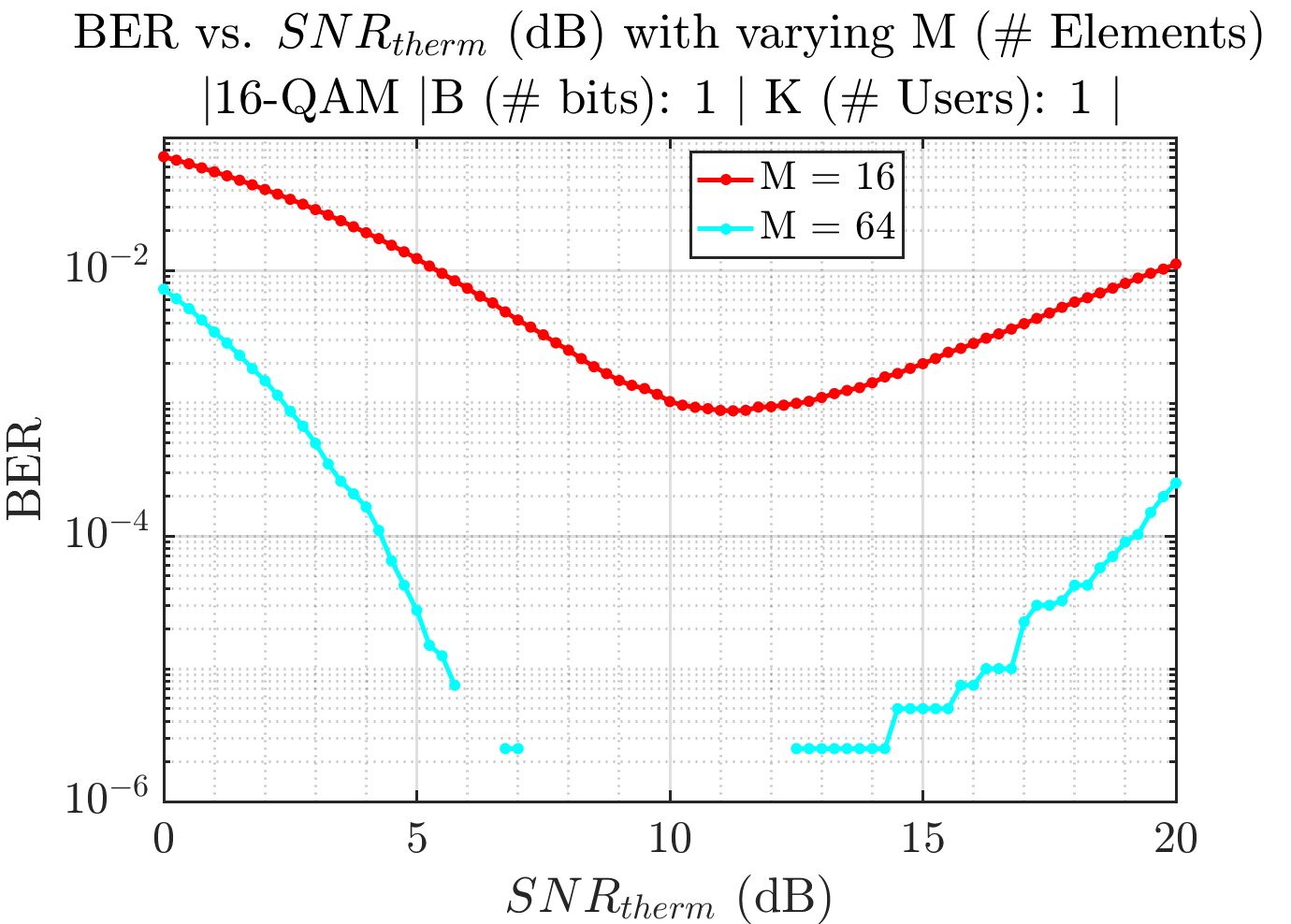}
    \caption{BER vs. $SNR_{therm}$ and SIR for varying M, B = 1}
    \label{fig:ChModeling2_BERvsSNRvsM_onebit}
\end{figure}

Nevertheless, it has been shown that one-bit ADCs can indeed support 16-QAM modulation \cite{studer}. This result was put to the test by the behavioral beamforming model presented here. Fig. \ref{fig:ChModeling2_BERvsSNRvsM_onebit} shows that, as expected, beamformers with one-bit ADCs can support 16-QAM modulation if the array is sufficiently large. In the case of a 16-element array, a bit-error rate close to $10^{-3}$ is achievable. Meanwhile, a 64-element array can reach BERs better than $10^{-5}$. However, as has been previously published this is only the case for SNR values that are neither too large or too small \cite{studer}. 

\subsection{Effects of User Count on System Performance}

In addition to determining system specifications for ADC resolution and minimum SIR, the behavioral model described here has been used to analyze the effects of user count on digital beamformer system performance. More specifically, this section explores the effects of user count on the array's quantization noise, since 1) we expect that the thermal noise is independent of user count and 2) the targeted digital beamforming architectures have low-resolution ADCs and, thus, significant quantization noise levels. 

\begin{figure}
    \centering
    \includegraphics[width=1\linewidth]{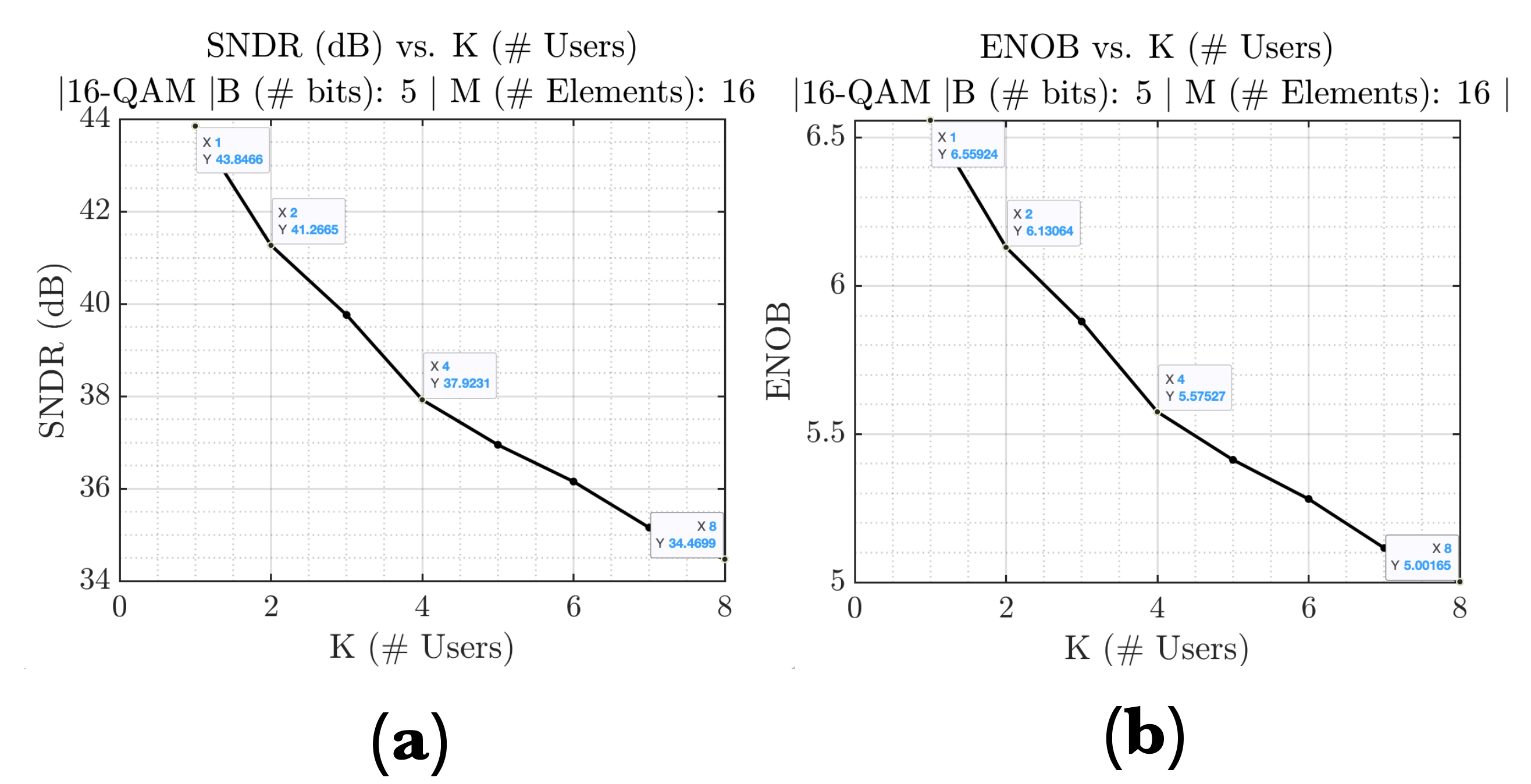}
    \caption{a) SNDR and b) ENOB vs. K, M = 16, B = 5, Zero-force Beamforming}
    \label{fig:ChModeling2_SNDRandENOBvsK_M16_ZF}
\end{figure}

\begin{figure}
    \centering
    \includegraphics[width=0.9\linewidth]{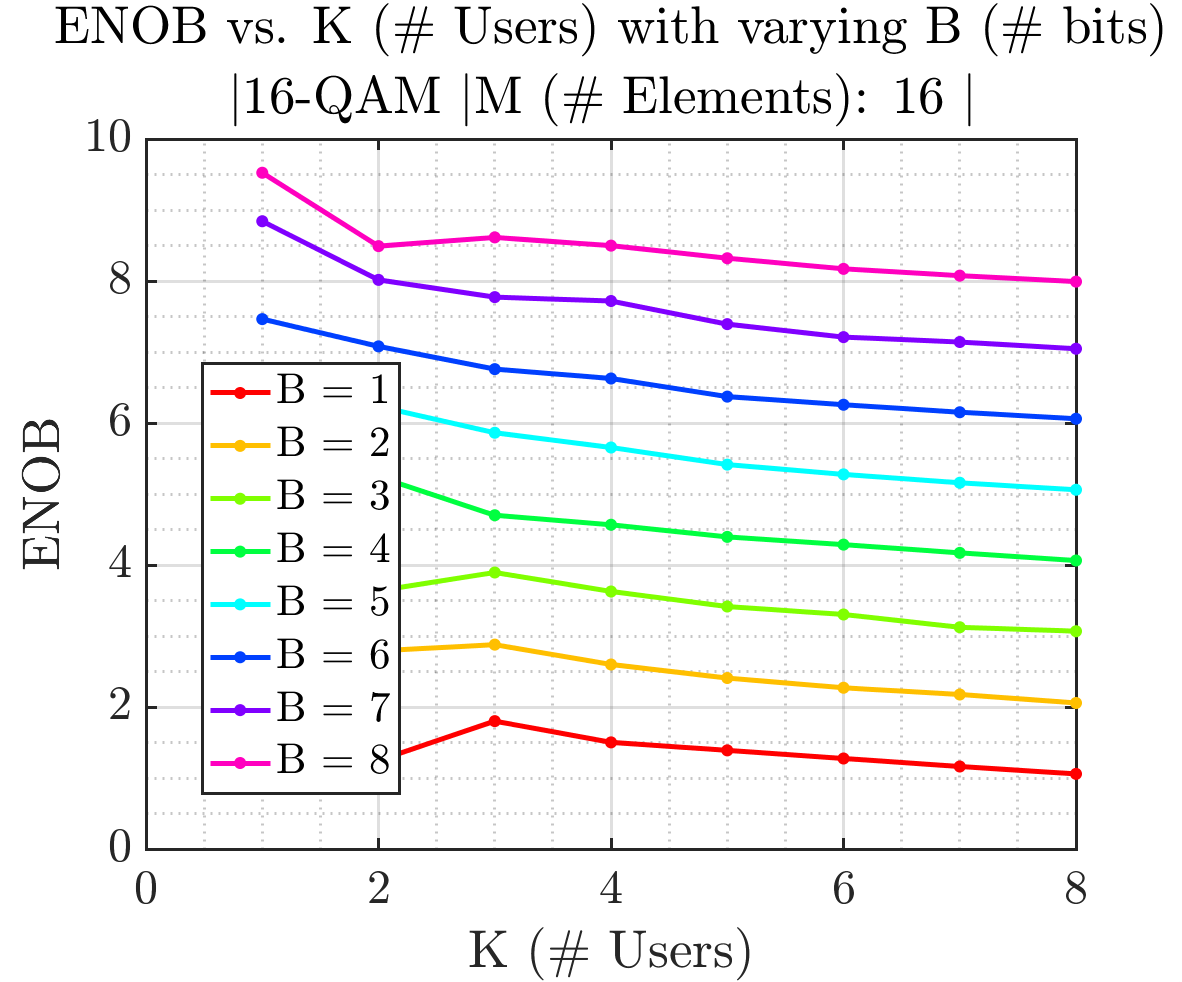}
    \caption{ENOB vs. K for varying B, M = 16, Zero-force Beamforming}
    \label{fig:ChModeling2_ENOBvsK_VaryB_M16_ZF}
\end{figure}

Fig \ref{fig:ChModeling2_SNDRandENOBvsK_M16_ZF} shows that, in the absence of an interferer, the SQNR worsens by approximately 3 dB as the number of users doubles. Equivalently, as the number of user doubles, ENOB degrades by approximately 0.5 bits. This trend can be observed for various ADC resolutions as shown in Fig. \ref{fig:ChModeling2_ENOBvsK_VaryB_M16_ZF}. 

Note that in this model, all users transmit at equal power levels and are equidistant from the receiver. Noting that the ADC full-scale is equally shared by each user ($P_{FS} = KP_{sig}$), Eqn.  \ref{eqn:SQNRwithFS} can be rewritten as follows:

\begin{equation}
    SQNR_{dB} = SQNR_{nom} - 10log_{10}(\frac{KP_{sig}}{P_{sig}})
\end{equation}

\begin{equation}
    SQNR_{dB} = SQNR_{nom} - 10log_{10}(K)
\end{equation}

\begin{figure}
    \centering
    \includegraphics[width=1\linewidth]{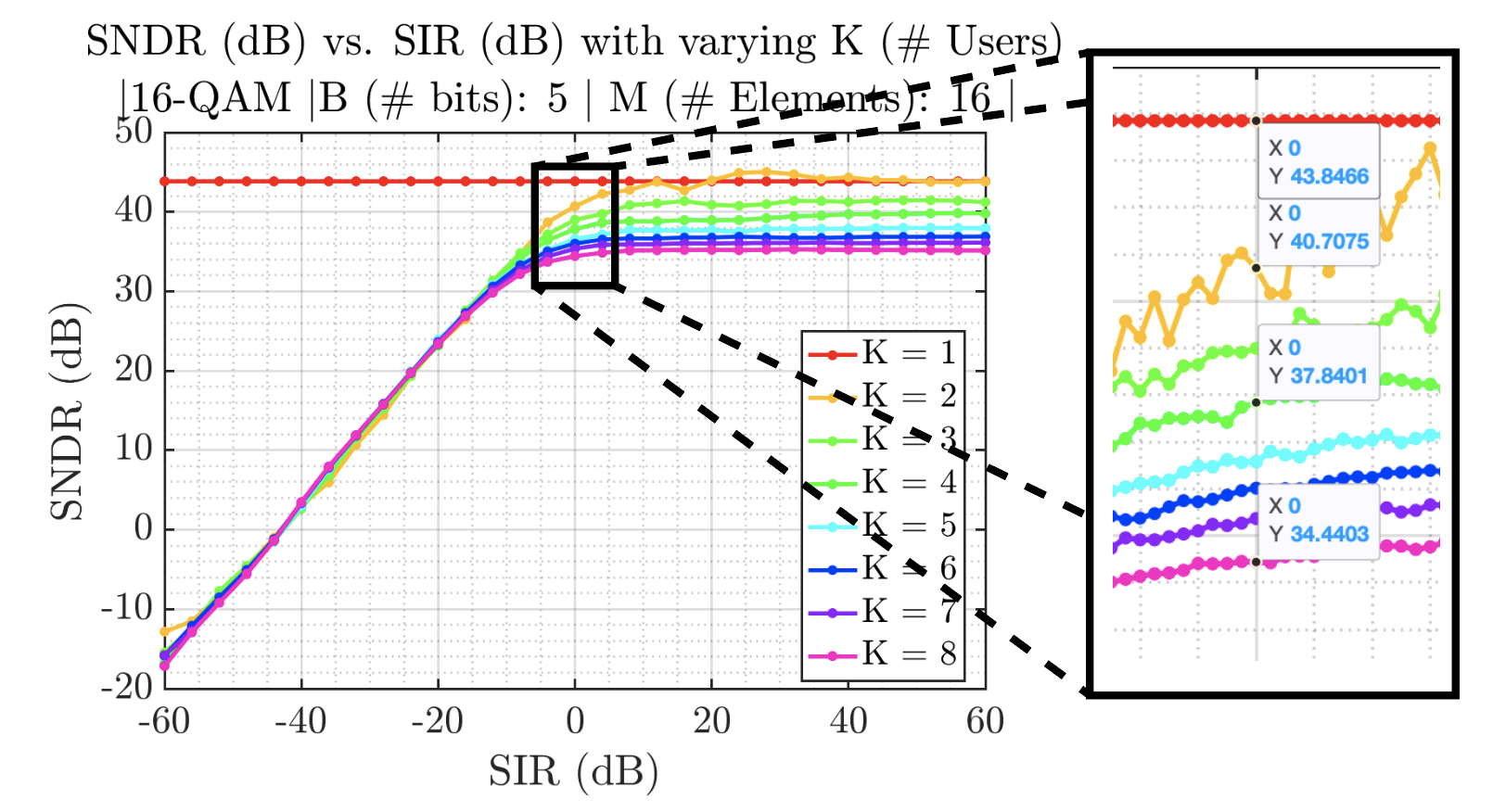}
    \caption{SNDR vs. SIR for with varying K, B = 5, M = 16, Zero-force Beamforming}
    \label{fig:ChModeling2_SNDRvsSIR_M16ZF_VaryK_Zoom}
\end{figure}

Recall that $SQNR_{nom}$ is nominally equal to $6.02B - 1.76$, though this varies with modulation scheme as previously described. Further, the effects of user count in the presence of a strong interferer is of interest since $SIR_{min}$ occurs when the degraded quantization noise dominates as has previously been shown. Fig. \ref{fig:ChModeling2_SNDRvsSIR_M16ZF_VaryK_Zoom} shows the SNDR vs. SIR for varying K. This result is similar to that shown in Fig. \ref{fig:SNDRvsSIR_1_16_CBF_ZF}c) where the ADC resolution is varied instead of the number of users. In contrast, however, Fig. \ref{fig:ChModeling2_SNDRvsSIR_M16ZF_VaryK_Zoom} shows that the maximum tolerable interferer strength is relatively equal across user count.

Since $K-1$ users transmit at nominal power and one interferer transmits at a power level depending on SIR, Eqn. \ref{eqn:SQNRwithFS} can again be rewritten as follows: 

\begin{equation}
    SQNR_{dB} = SQNR_{nom} - 10log_{10}(\frac{(K-1)P_{sig}+P_{interf}}{P_{sig}})
\end{equation}

\begin{equation}
    SQNR_{dB} = SQNR_{nom} - 10log_{10}((K-1) + \frac{1}{SIR})
    \label{eqn:SQNRvsKandSIR}
\end{equation}

Fig. \ref{fig:ChModeling2_SNDRvsSIR_M16ZF_VaryK_Zoom} visualizes the behavior described by Eqn. \ref{eqn:SQNRvsKandSIR}. When the interferer is as strong or weaker than the other users (SIR $\geq$ 0), the SQNR depends on K, worsening by approximately 3 dB as the number of users doubles. Meanwhile, when the interferer is strong, $SQNR_{dB} = SQNR_{nom} + SIR_{dB}$, so the SNDR tracks the SIR. As a result, in the presence of strong interference the SQNR is not a strong function of the number of users. Thus, $SIR_{min}$, as defined in this paper, does not vary strongly with the number of users either.

\section{Conclusions}
\label{section:ChDBFConclusions}

Digital beamforming forms the foundation for massive MIMO in next-generation wireless communications. At its core, digital beamforming architectures provide key benefits such as faster beam search, interference nulling via zero-force beamforming, higher spectral capacity, and more increased flexibility. However, they generally tradeoff power consumption due to the large number of ADCs in such systems. 

The behavioral model presented in this paper aims to deepen understanding of such digital beamforming systems to enable system designers to make optimizations. The results presented in this paper primarily center on implementations with low-resolution ADCs and, thus, focus on the effects of system parameters, including interferer strength, on  quantization noise.

This paper's contributions are summarized below:
\begin{itemize}
    \item An open-source end-to-end digital beamforming system model is presented and described in detail for other members in the community to use in their own investigations. A user interface is wrapped around the model to enable users to recreate figures presented in this paper and to allow users to tune various parameters to match their system of interest.
    \item An equation-based model for SNDR and SQNR degradation due to strong interference is proposed and verified with the behavioral model.
    \item A new heuristic, $SIR_{min}$ is introduced to determine the maximum tolerable interferer strength for the 16-QAM communication link of interest.
    \item An analysis of digital beamformers with low-resolution ADCs in the presence of interference is presented to verify the viability of 3-4 bits in modestly sized arrays.
    \item A case study on one-bit ADCs for digital beamformers is presented to verify the necessary conditions of operation for such a system.
    \item A study on the effects of user count on digital beamformers with and without interference is presented. 
\end{itemize}

Additionally, below is a list of potential future improvements and directions:
\begin{itemize}
    \item Add support for analog and hybrid beamforming to the model to draw direct comparisons between architectures
    \item Add support for more realistic MIMO communication channels, including non-frequency-flat fading effects due to multi-path. 
    \item Add matched filtering at the RX and TX.
    \item Develop a heuristic for minimum required ADC resolution.
    \item Add support for other modulation schemes and include effects of RF non-linearities and channel mismatch.
\end{itemize}

The open-source nature of the work presented is intentional. The results presented here are a subset of the observations that can be made with such a model. The primary motivation for sharing this open-source model is to enable others to validate their own observations or hypotheses about these complex systems. Naturally, potential errors in the codebase may emerge as users find edge cases. However, some of the results presented in this paper first appeared to be errors - it is important to remain skeptical of both our instinctive intuition for such systems and for potential codebase errors. Nevertheless, the hope is that this model's codebase and the results presented serve as a starting point for furthering our understanding of large communication systems.

\bibliographystyle{ieeetr} 
\bibliography{bibtex/bib/DBFModelPaper}

\appendices
\section{GitHub Repository and Usage Instructions}
\label{section:appendix}
\subsection{GitHub Repository}
The public GitHub repository is available at the following link:
\url{https://github.com/joseguaj1/OSBehavioralDBFModel}

\subsection{Required MATLAB Toolboxes}
As of writing, the model is used on MATLAB version R2024b. The model makes use of a few necessary MATLAB Toolboxes that must be installed: 
\begin{itemize}
    \item  MATLAB R2024b
    \item Signal Processing Toolbox
    \item Communications Toolbox
    \item Curve Fitting Toolbox
\end{itemize}

\subsection{Graphical User Interface (GUI) Overview}
\begin{figure}
    \centering
    \includegraphics[width=1\linewidth]{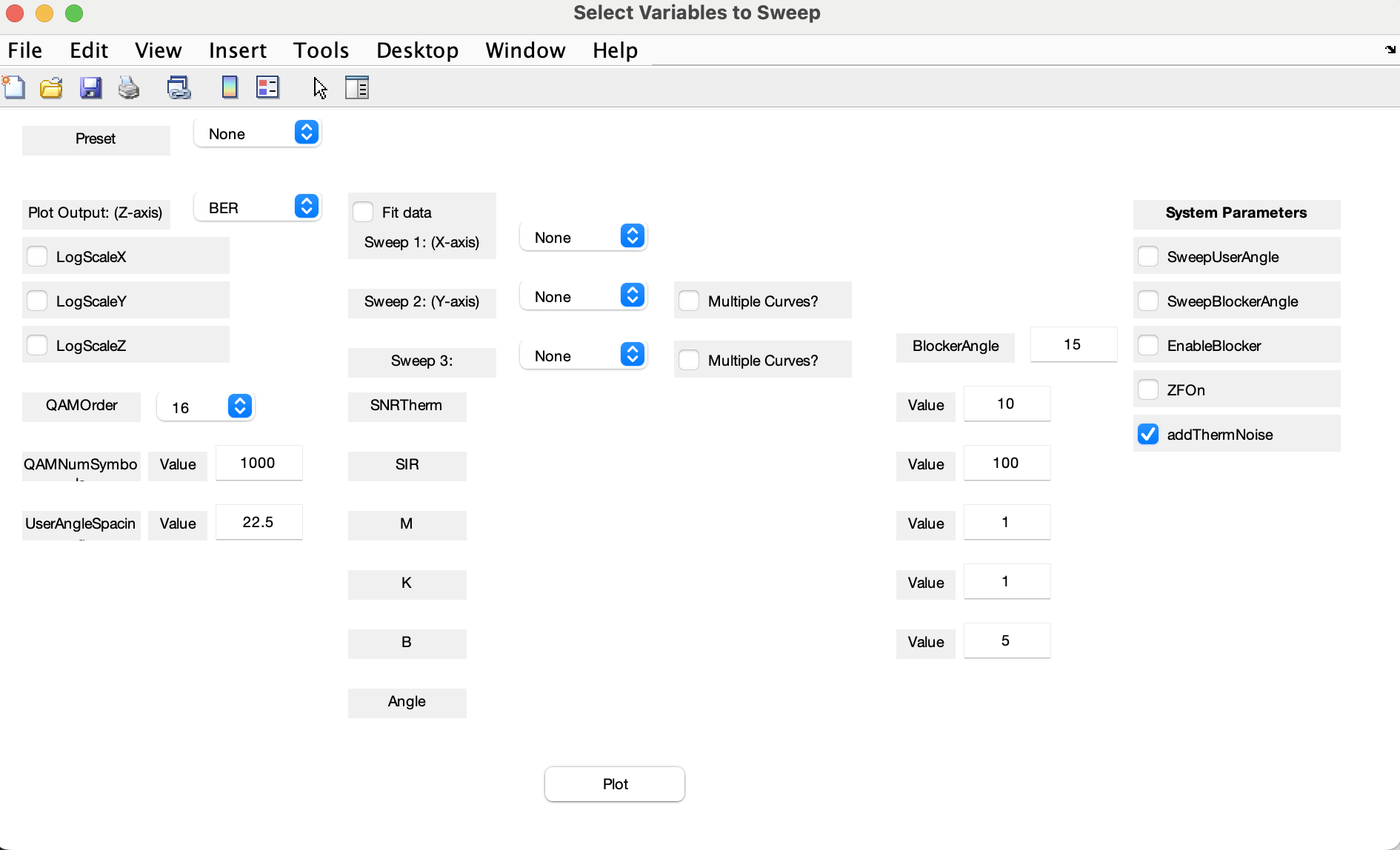}
    \caption{GUI to Interact with Behavioral Model}
    \label{fig:ChModeling3_GUI}
\end{figure}

Fig. \ref{fig:ChModeling3_GUI} shows the GUI. There are several key parts of the GUI to note. 
\subsubsection{Presets and Outputs}
\begin{figure}
    \centering
    \includegraphics[width=1\linewidth]{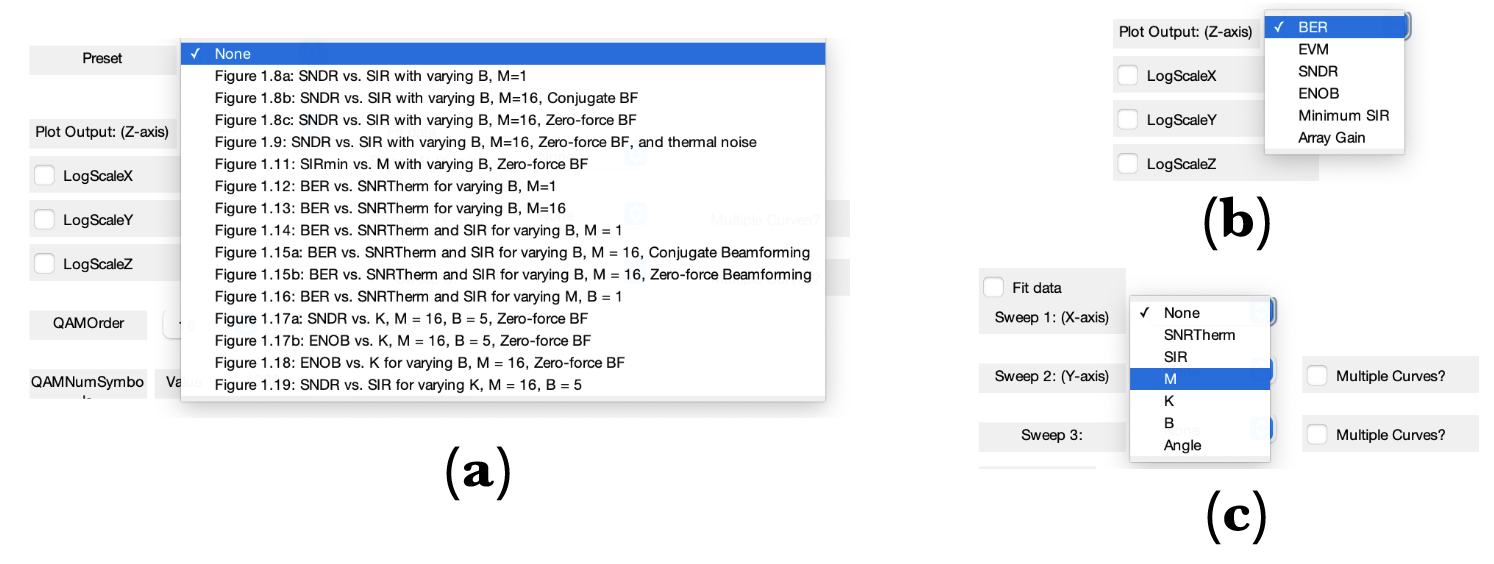}
    \caption{GUI Dropdown Menus for a) Presets b) Outputs c) Sweepable variables}
    \label{fig:ChModeling3_GUI_Dropdowns}
\end{figure}

The presets dropdown menu is shown in Fig. \ref{fig:ChModeling3_GUI_Dropdowns}a) and can be used to replicate any of the results presented in this paper. Additionally, the plot output dropdown menu can be used to select the output of interest, as shown in Fig. \ref{fig:ChModeling3_GUI_Dropdowns}b). 

\subsubsection{Sweepable Variables}
    Fig. \ref{fig:ChModeling3_GUI_Dropdowns}c) shows the dropdown menu listing the variables that can be swept. Note that up to three variables can be swept and the output plot will match the necessary number of dimensions. Note that the 'Multiple Curves' checkboxes may be used to determine whether a second or third sweep variable will appear as a plot axis or as a set of curves. Additionally, when a sweep variable is selected, the user will be able to input the start, end and step values for that variable. If the variable is not being swept, it takes the default value from the box labeled 'Value'. 
    
\subsubsection{Boolean Variables}
    There are two sets of boolean variables that can be changed with the GUI. Firstly, the LogScale variables found on the left side of the GUI can be enabled in order to plot certain axes with a logarithmic scale. This is useful, for example, when plotting bit-error rates. There is a second set of boolean variables, labeled System Parameters, on the right side of the GUI that can also be enabled or disabled as needed.

\subsubsection{Non-sweepable Variables}
    Lastly, there are a few variables that can be tuned but not swept. Namely, these are QAMNumSymbols, UserAngleSpacing and BlockerAngle. Note that the only supported modulation scheme is 16-QAM.

\subsubsection{User Placement}

\begin{figure}
    \centering
    \includegraphics[width=1\linewidth]{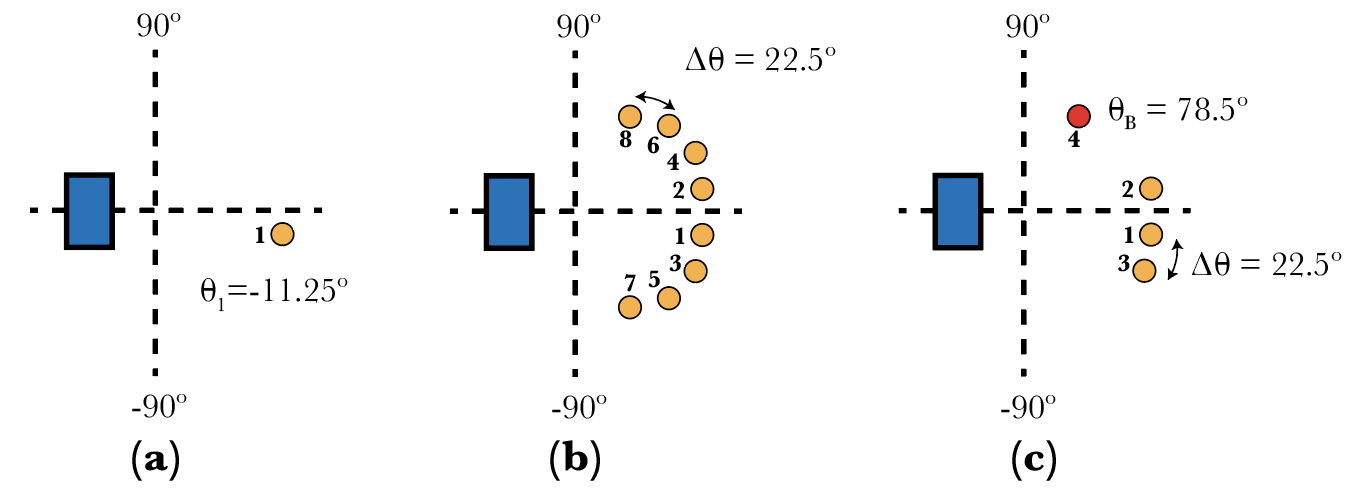}
    \caption{User Placement when $BlockerAngle  = 78.5^{\circ}$ and $UserAngleSpacing =22.5^{\circ}$ for a) $M = 1$ b) $M=8$ c)  $M = 4$ with Blocker Enabled}
    \label{fig:ChModeling3_UserPlacement}
\end{figure}

Most of the tunable parameters are straightforward. However, there are two tuning parameters that must be tuned carefully:

\begin{itemize}
    \item BlockerAngle
    \item UserAngleSpacing
\end{itemize}

BlockerAngle and UserAngleSpacing must be chosen carefully to ensure that the users are placed as desired. Fig. \ref{fig:ChModeling3_UserPlacement} shows how the model places users depending on BlockerAngle and UserAngleSpacing. For the user placements shown in Fig. \ref{fig:ChModeling3_UserPlacement}, $BlockerAngle  = 78.5^{\circ}$ and $UserAngleSpacing =22.5^{\circ}$. We note that users are always placed around but not at zero degrees (broadside). Fig. \ref{fig:ChModeling3_UserPlacement}a) shows that for a single user, $\theta_1 = -UserAngleSpacing/2$. As the number of users are increased, each one is placed an angular distance of UserAngleSpacing apart from the first user. Note the order in which users are placed in the figure. Fig. \ref{fig:ChModeling3_UserPlacement}b) shows the arrangement when $K = 8$. Lastly, Fig. \ref{fig:ChModeling3_UserPlacement}c) shows the arrangement when $K=4$ and the interferer, shown in red, is enabled. 

\end{document}